\documentclass[preprint,11pt]{JHEP3} 
\JHEPspecialurl{http://jhep.sissa.it/JOURNAL/JHEP3.tar.gz}
\usepackage{epsfig,multicol,amsmath,mathrsfs}

\usepackage{graphicx}
\usepackage{epsfig,multicol}
\usepackage{pifont}
\usepackage{relsize}
\def\beq{\begin{equation}}
\def\eeq{\end{equation}}
\def\bea{\begin{eqnarray}}
\def\eea{\end{eqnarray}}

\def\eq#1{{Eq.~(\ref{#1})}}
\def\fig#1{{Fig.~\ref{#1}}}
\def\sec#1{{section~\ref{#1}}}

\newcommand{\bas}{\bar{\alpha}_s}
\newcommand{\as}{\alpha_s}

\newcommand{\prm}{^{\,\prime}}
\newcommand{\dpr}{^{\,\prime\prime}}
\newcommand{\He}{=\hspace{0.3cm}}
\newcommand{\he}{\hspace{0.3cm}=\hspace{0.3cm}}

\newcommand{\Lb}{\left(}
\newcommand{\Rb}{\right)}

\parskip=\itemsep               

\newcommand{\nn}{\nonumber}
\newcommand{\D}{\partial}

\newcommand{\h}{\frac{1}{2}}

\newcommand{\ga}{\gamma}
\newcommand{\de}{\delta}
\newcommand{\De}{\Delta}

\newcommand{\la}{\lambda}

\newcommand{\si}{\sigma}

\newcommand{\Ga}{\Gamma}
\newcommand{\om}{\omega}
\newcommand{\Om}{\Omega}
\newcommand{\La}{\Lambda}

\renewcommand{\theequation}{\thesection.\arabic{equation}}

\newcommand{\lab}{\label}
\newcommand{\f}{\frac}

\newcommand{\lf}{\mathlarger{\f{1}{4}}}
\newcommand{\lh}{\mathlarger{\h}}
\newcommand{\lhh}{\mathlarger{\f{3}{2}}}

\newcommand{\ml}{\mathlarger}
\newcommand{\bl}{\biggl(}
\newcommand{\br}{\biggr)}

\newcommand{\Lint}{\ml{\ml{\int}}}

\title{\LARGE \bf The BFKL Pomeron calculus: summing enhanced diagrams}
\author{\Large 
E.\, Levin${}^{a, b}$ \thanks{Email: leving@post.tau.ac.il., eugeny.levin@usm.cl}\,\,\,and \,\, J.\,Miller${}^{a, c}$
\thanks{Email:  jeremy.miller@ist.utl.pt}
\\
${}^a$ \, Department of Particle Physics, School of Physics and Astronomy,
Tel Aviv University, Tel Aviv, 69978, Israel\\
${}^b$\, Departamento de F\'\i sica,
Universidad T$\acute{e}$cnica Federico Santa Mar\'\i a, and
Centro Cientifico-Tecnol$\acute{o}$gico de Valpara\'\i so,
Casilla 110-V,  Valparaiso, Chile\\ 
${}^c$\,CENTRA, Departamento de F\'\i sica, Instituto Superior T\'ecnico (IST), Av. Rovisco Pais, 1049-001 Lisboa, Portugal
}

\abstract{ The goal of this paper is to sum over a class of enhanced diagrams,
and derive a new Pomeron Green function.  It is found that this sum gives the Pomeron contribution to the scattering amplitude 
that decreases with energy. In other words, we found that the total cross section of two colourless dipoles of small but equal sizes, falls down at high energies.
 }
\keywords{colourless, triple Pomeron vertex, BFKL Pomeron calculus, high energy behaviour of the amplitude}
\preprint{ \today}

\begin{document}

\section{Introduction}

High energy QCD has reached a mature stage of  development, for dilute-dense scattering
(for example for DIS with nuclei) \cite{GLR,MUQI,MV,JIMWLK,B,K}. However,
 for  dilute-dilute scattering at high energy, (for example the scattering of two virtual photons with large but almost equal virtualities),
despite the great deal of effort by experts in the field (see for example Refs.
 \cite{MSW,LL,IT,KOLU,HIMST,BRAUN,AKLP,LMP,KLM} ), we are still waiting for the desired breakthrough.
In this paper we address the problem of the scattering amplitude for the dilute - dilute system, using the BFKL Pomeron calculus \cite{BFKL,LIPREV,GLR,MUQI,BRAUN,BFKL,BART}.
The BFKL Pomeron calculus is
elegantly formulated 
in terms of the functional integral \cite{BRAUN}.  Namely,

\begin{equation} \label{BFKLFI}
Z[\Phi, \Phi^+]\,\,=\,\,\int \,\,D \Phi\,D\Phi^+\,e^S \,\,\,\hspace{0.5cm}\mbox{with}\hspace{0.5cm}\,S \,=\,S_0
\,+\,S_I\,+\,S_E
\end{equation}

where $S_0$ describes free Pomerons, $S_I$ corresponds to their mutual interaction
while $S_E$ relates to the interaction with the external sources (target and
projectile). $S_0$ has a simple expression:

\begin{equation} \label{S0}
S_0\,=\,\int\,d Y \,d Y'\,d^2 x_1\, d^2 x_2\,d^2 x'_1\, d^2 x'_2\,
\Phi^+(x_1,x_2;Y)\,
G^{-1}(x_1,x_2;Y|x'_1,x'_2;Y')\,\Phi(x'_1,x'_2;Y')
\end{equation}

while

\beq \label{SI}
S_I\,=\,\frac{2\,\pi \bas^2}{N_c}\,\int \,d Y'\,\int
\,\frac{d^2 x_1\,d^2 x_2\,d^2 x_3}{x^2_{12}\,x^2_{23}\,x^2_{13}}
\Big\{ \left( L_{12}\Phi(x_1,x_2;Y')\,\right)\,
\Phi^+(x_1,x_3;Y')\,\Phi^+(x_3,x_2;Y')\,\,+\,\,h.c. \Big\} \eeq

where $h.c.$ denotes the Hermitian conjugate.
$S_E$ describes the interaction with the scattering particles (two small colliding dipoles in our case),
 but we do not need the explicit expression for this term in our paper.
 $Y$ is the rapidity of the dipole. For dipole-dipole scattering  $Y = \ln s $  in the leading log approximation where $s$ is the energy of colliding dipoles.
 \eq{BFKLFI} is written in the leading log approximation. The following notation, $\bas = N_c \as/\pi$ will be assumed
throughout.
 $\Phi$ and $\Phi^+$  relate to the BFKL Pomeron and $G(x_1,x_2;Y|x'_1,x'_2;Y')$ is the Green function of the BFKL Pomeron which takes the form:

\begin{equation}
G^{-1}(x_1,x_2;Y| x'_1,x'_2;Y')\he p^2_1\,p^2_2\,\left(
\frac{\partial}{\partial Y} + H \right) \he \left(
\frac{\partial}{\partial Y} + H^+ \right)\,p^2_1\,p^2_2 \label{G1}
\end{equation} 

with

\begin{equation}
H f(x_1,x_2;Y) \,\,= \,\,\frac{\bas}{2
\pi}\,\int\,d^2 x_3\, K\Lb x_1, x_2|  x_3\Rb\,\left\{
f(x_1,x_2;Y)\,-\,f(x_1,x_3;Y)\,-\,f(x_3,x_2;Y) \right\} \label{H}
\end{equation}

where

\beq \label{K}
K\Lb x_1, x_2|  x_3\Rb\he \frac{
x^2_{12}}{x^2_{23}\,x^2_{13}}\hspace{2cm} L_{12} \he x^4_{12} \,p^2_1\,p^2_2 \hspace{2cm}p_k\he -i \nabla_{x_k}\hspace{0.5cm}\Lb k\,=\,1,2\Rb
\eeq

The exact Green function for the Pomeron, which is the goal of this paper to derive, is equal to:

\beq \label{PHIG}
G(x_1,x_2;Y| x'_1,x'_2;Y')\,\,=\,-\frac{\int D \Phi\,D\Phi^+    \Phi\Lb x_1, x_2; Y \Rb \,\Phi^+\Lb x'_1.x'_2; Y'\Rb\,e^{S[\Phi,\Phi^+]}}{ \int D \Phi\,D\Phi^+  \,e^{S[\Phi,\Phi^+]}}
\eeq
and the bare (initial)  BFKL Pomeron Green function\cite{LIPREV} is determined by \eq{PHIG},
where  only the term $S_0$ is included in $S[\Phi,\Phi^+]$.

 It is worthwhile mentioning that due to conformal invariance of the BFKL Pomeron calculus, the form of the triple Pomeron vertex is known  unambiguously,
  and coincides with the direct calculations found in ref. \cite{BART}\footnote{For the sake of completeness in this presentation, we would like to mention that  there is still a discussion in the literature, about
 whether or not the BFKL Pomeron calculus in the form of  Eq. (1.1),  correctly takes into account the reggeized 
gluons. However, Eq. (1.1) reproduces the non-linear Balitsky-Kovchegov equation, including the term for the gluon reggeization.
  As far as we know, no other examples have been suggested, where  Eq.(1.1) provides an incorrect result
 due to problems with gluon reggeization."}
The simplicity of \eq{BFKLFI} is thanks to the key assumption, 
that the triple Pomeron vertex is the only essential vertex for Pomeron interactions. 
 In other words. in \eq{BFKLFI} we neglect all other local vertices (for example, the vertex for the transition of one Pomeron to three and so on).
 We have no rigorous proof for this conjecture.
 There even exist arguments that the four Pomeron vertex should also be included \cite{KLP}.
 It should noted also,  that the BFKL Pomeron calculus given by \eq{BFKLFI}, is formulated in the leading $1/N_c$ approximation,
 where $N_c$ is the number of colours. It is known that
 the BFKL Pomeron calculus cannot be a correct approximation for the scattering amplitude in the next-to-leading order in $1/N_c$ approach, 
due to the fact that $2 n $- gluon states in the $t$-channel give a larger intercept than $n$ BFKL Pomerons \cite{MPOM}.
  Nevertheless we consider \eq{BFKLFI} to be a good first approximation, for the dilute-dilute system of scattering. 
 We will return to the discussion of all these problems in the conclusion, where we show that they are not important for the solution derived in this paper.

 The goal of this paper is to calculate the class of enhanced diagrams. The simplest examples of these diagrams are  shown in \fig{f1loop} and \fig{fnloop}. In other words we are going to sum BFKL Pomeron loops in this paper.
These diagrams lead to a new Green function of the BFKL Pomeron  (the term $S_0$ in \eq{BFKLFI} ), while the vertices of the interaction of the new  dressed Pomeron
 remain the same as they appear in $S_I$.  It is worthwhile mentioning that if the Green function of the dressed Pomeron, will be such that the Pomeron contribution will lead 
to a  decrease  with energy, then the problem would be solved without needing to consider the interaction of the dressed Pomerons.
However, if this is not the case and the dressed Pomeron still increases with energy,  then the
interaction of the dressed Pomeron needs to be included.
On the other hand, for the scattering of two dipoles with small sizes, within a wide range of energy the enhanced diagrams dominate,
 since the interactions of the Pomeron with the target and the projectile are small {in the leading $1/N_c$-approximation.

This paper is organized in the following way.
 In the next section the formulae for the triple Pomeron vertex and the Pomeron Green function are introduced.
 Using these ingredients
 we calculate the sum over the class of enhanced diagrams shown in \fig{f1loop} - \fig{fnloop}, namely one-Pomeron loops in series, in $\Lb\om,\nu\Rb$
representation. 
This is done in a step-by-step way,
in order to introduce the reader to the method of integration and assumptions used in this treatment, which will be used
later on for more complicated diagrams.

The third section is the main body of this paper.
Using the techniques developed in \sec{sloops},  we extend  this approach to the sum over all enhanced diagrams,
 using two principle selection rules. First, we are searching only for the contribution to the vertices that are singular in $\nu$.
   Second, we assume that the contribution of the Pomeron loops in effective vertices, for multi-Pomeron production are negligibly small.
This approximation is closely related to the Mueller-Patel-Salam-Iancu 
approach \cite{MPSI}, but in this paper  this strategy is formulated in $(\om, \nu)$-representation.
In section 3.1 the equations for the effective  multi-Pomeron vertices are derived and solved.
 In section 3.2 we present the results of the summation, for the Green function of the dressed Pomeron. It turns out that the exchange of the dressed Pomeron, leads to the total cross section for dipole-dipole scattering that decreases with energy.
 This result is reminiscent of the Pomeron Green function  in   $1+1$ dimensional Pomeron calculus, \cite{AAJ,ALMC,CLR,CI,BOMU} 
which also decreases due to the contribution of Pomeron loops.
 In the conclusion section, we summarize our results, and discuss the current stage of development of the BFKL Pomeron calculus.
 The appendix provides detailed information about the ingredients of the BFKL Pomeron calculus, including the bare Pomeron Green function and the triple Pomeron vertex.

\section{Summing simple Pomeron loops}\lab{sloops}

In this section we sum the set of enhanced diagrams shown in \fig{f1loop} - \fig{fnloop}, namely
one-Pomeron loops in series.
The sum over all diagrams of this type
 will teach us the main characteristic features of the enhanced diagrams, which we will use for developing an approach for summing over a more general class of  diagrams.

\subsection{Bare Pomeron}

\DOUBLEFIGURE[t]{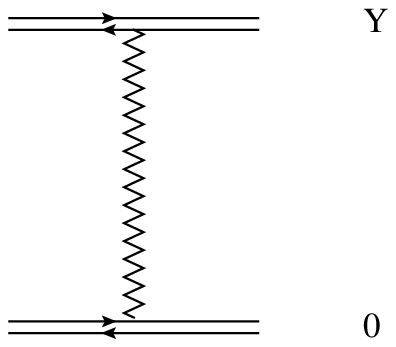,width=50mm,height=45mm}{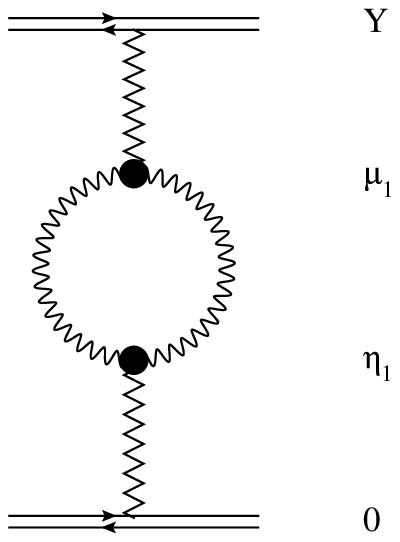,width=45mm,height=45mm}
{The basic diagram with the exchange of one Pomeron in the $t$-channel.
\label{f1pomeron}}{The diagram with one Pomeron loop. \label{f1loop} }

The most basic diagram is the exchange of one Pomeron in the $t$-channel without any loops. The expression for this diagram 
which is shown in 
\fig{f1pomeron}
is given  by the following expression  in $\om$ representation \cite{1.Navelet:2002zz,Navelet:1998yv,2.Navelet:2002zz,Bialas:1997xp,Kozlov:2004sh}:

\bea\lab{1pomomega}
&&G_{0}\Lb\om,\nu\Rb\he\,\f{g\Lb\nu\Rb}{\om-\om\Lb\nu\Rb}\,\,\,\,\,\,\,\,\,\,\,\,\,
\mbox{where}\hspace{0.5cm} g\Lb\nu\Rb\he\f{\nu^2}{\biggl(\nu^2+\lf\biggr)^2}\lab{gnu}\eea

The following inverse Mellin transform allows us
to pass to $Y$ representation:

\bea&&
G_{0}\Lb Y,\nu\Rb\he\f{1}{2\pi i}\Lint^{a+i\infty}_{a-i\infty}d\om \,\,\,e^{\om\,Y}G_{(0)}\Lb\om,\nu\Rb
\he\,e^{\,\om\Lb\nu\Rb\,Y}\,g\Lb\nu\Rb\,\lab{1pomY}\eea

The contour of integration is placed to the right of all singularities of $G_0\Lb \omega,\nu\Rb$.

\subsection{One loop}
\lab{s1loop}

The first correction to the basic diagram is the diagram with one loop
shown below in \fig{f1loop}.
The amplitude for the one-loop diagram takes the following form in $\om$ representation \cite{BRAUN}:

\bea
&&
A_{(1)}\Lb\om,\nu\Rb\he G_0\Lb\om,\nu\Rb\Lint^\infty_{-\infty}\!\!d\nu\prm\,\, m\Lb\om,\nu,\nu\prm\Rb G_0\Lb\om,\nu\prm\Rb \lab{1loopomega}\eea

where $m\Lb\nu,\nu\prm,\om\Rb$ is the Pomeron self mass, given by the following formula:

\bea
&&m\Lb\nu,\nu\prm,\om\Rb\he\,\Lint^\infty_{-\infty}d\nu_1\,\Lint^\infty_{-\infty}d\nu_2\,g\Lb\nu_1\Rb\,g\Lb\nu_2\Rb
\ml{\f{ \Ga\biggl(\nu\,\begin{vmatrix}\,\nu_1,\nu_2\biggr)  \Ga\biggl(\nu\prm\,\end{vmatrix}\,\nu_1,\nu_2\biggr) }{\bl\,\om-\om\Lb\nu_1\Rb-\om\Lb\nu_2\Rb\,\br}}\hspace{1cm}
\lab{definitionSigma}\eea

The triple Pomeron vertex has the following definition \cite{1.Korchemsky:1997fy}
\footnote{The coefficient in front of
 \eq{tpv} contains an extra factor of $1/N_c$ in accordance \eq{SI}, compared to the same coefficient that appears in
\cite{1.Korchemsky:1997fy}.}:

\bea
&&\Ga\Lb\nu\,\vert\,\nu_1,\nu_2\Rb\he\frac{16\,\bas^2}{N_c}\Lb \Ga_{\mbox{\footnotesize{planar}}}\Lb\nu\,\vert\,\nu_1,\nu_2\Rb
-\f{2\pi}{N_c^2}\,\, \Ga_{\mbox{\footnotesize{non-planar}}}\Lb\nu\,\vert\,\nu_1,\nu_2\Rb\Rb\lab{tpv}\\
\nn\\
&& \Ga_{\mbox{\footnotesize{planar}}}\Lb\nu\,\vert\,\nu_1,\nu_2\Rb
\he \Lb\lf+\nu^2\Rb^2\Om\Lb\nu\,\vert\,\nu_1,\nu_2\Rb\lab{planar}\\
\nn\\
&& \Ga_{\mbox{\footnotesize{nonplanar}}}\Lb\nu\,\vert\,\nu_1,\nu_2\Rb
\he \Lb\lf+\nu^2\Rb^2\La\Lb\nu\,\vert\,\nu_1,\nu_2\Rb\bl
\chi\Lb\nu\Rb-\chi\Lb\nu_1\Rb-\chi\Lb\nu_2\Rb\br\lab{nonplanar}
\eea
where the functions $\chi\Lb\nu\Rb$\,, $\,\,\Om\Lb\nu\,\vert\,\nu_1,\nu_2\Rb$ and $\La\Lb\nu\,\vert\,\nu_1,\nu_2\Rb$ are defined explicitly in Eqs. (\ref{chi}),  
(\ref{thesumofomegas}) and (\ref{lanu}) respectively. $ \Ga_{\mbox{\footnotesize{planar}}}$ and $ \Ga_{\mbox{\footnotesize{nonplanar}}}$ denote 
the two diagrams that contribute to the vertex, namely the planar and non-planar diagrams shown below in \fig{fvertex} (a) and \fig{fvertex} (b) 
(the diagrams in \fig{fvertex} are taken from ref. \cite{1.Korchemsky:1997fy}).

\FIGURE[h]{\begin{minipage}{150mm}
\centerline{\epsfig{file=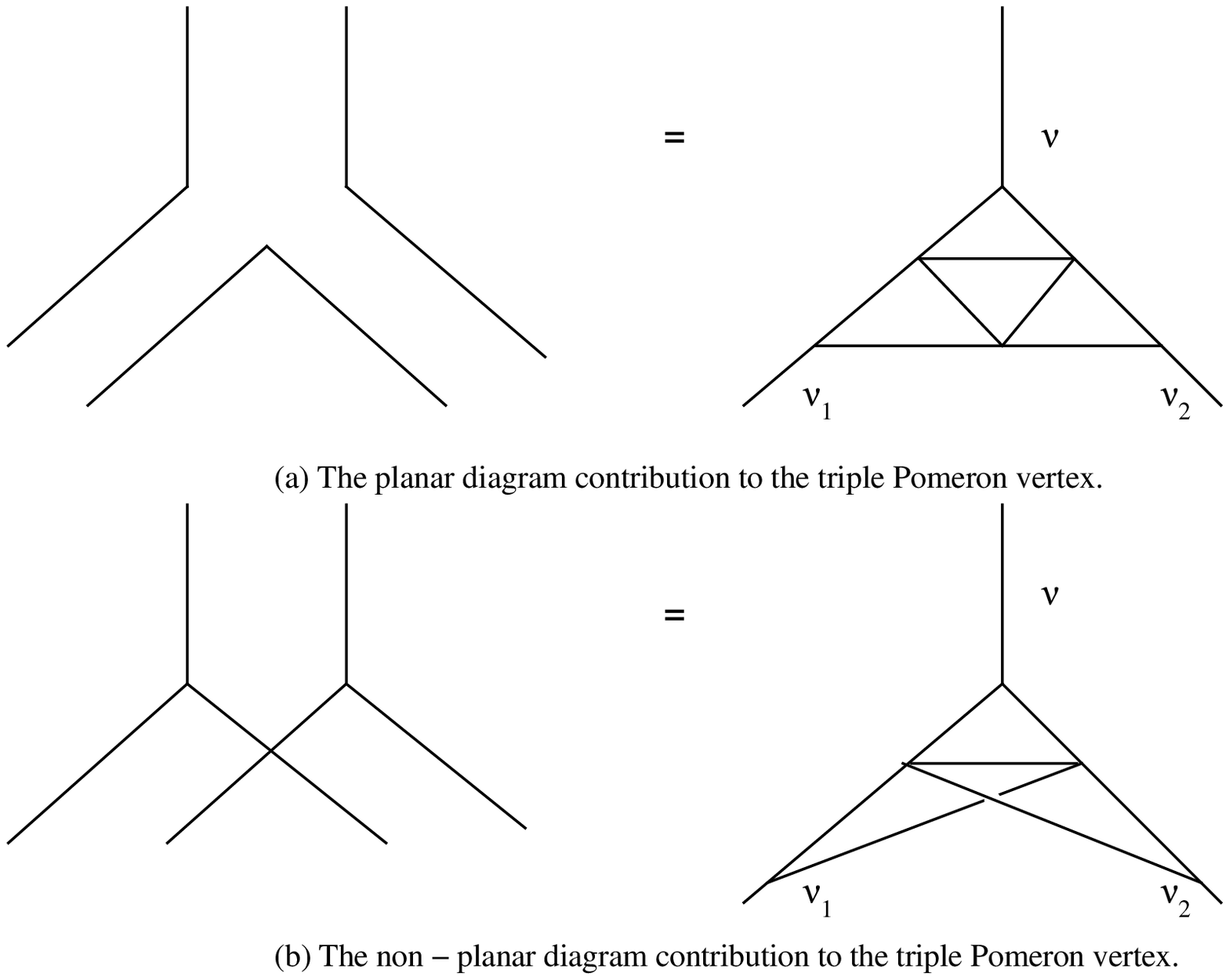,width=140mm}}
\end{minipage}
\caption{The two contributing diagrams to the triple Pomeron vertex.
} 
\label{fvertex}}

$\Ga\Lb\nu\,\vert\,\nu_1,\nu_2\Rb$ is the vertex at the top of the loop of \fig{f1loop}, whereby the Pomeron with  scaling dimension $\nu$ splits into two Pomerons that 
form the loop, with  scaling dimensions $\nu_1$ and $\nu_2$. The vertex at the bottom of the loop is labeled by $\Ga\Lb\nu\prm\,\vert\,\nu_1,\nu_2\Rb$,
whereby the Pomerons with  scaling dimensions $\nu_1$ and $\nu_2$ recombine to one Pomeron with  scaling dimension $\nu\prm$.
Before proceeding to calculate $m\Lb\om,\nu,\nu\prm\Rb$, it is instructive to switch to the variables $\la$, $\si$ and $\De$ defined as:

\bea&&
\la\he i/2 - \nu\,;\hspace{1cm}\la\prm\he i/2 - \nu\prm\,;\hspace{1cm}
\si\he\nu_1+\nu_2\,;\hspace{1cm}
\De\he\nu_1-\nu_2\lab{depm}\eea

In tems of ($\la,\si,\De$) variables, then \eq{definitionSigma} reads:

\bea&&
m\Lb\om,\la,\la\prm\Rb\he
\Lint^\infty_{-\infty}d\si\Lint^\infty_{-\infty}d\De\,\, g\Lb\ml{\f{\si+\De}{2}}\Rb g\Lb\ml{\f{\si -\De}{2}}\Rb
\f{\Ga\bl\la\,\begin{vmatrix}\,\si,\De\br\Ga\bl\la\prm\,\end{vmatrix}\,\si,\De\br}{\Lb\,\om-\om\Lb\ml{\f{\si+\De}{2}}\Rb-\om\Lb\ml{\f{\si-\De}{2}}\Rb\,\Rb}\,\,\hspace{1cm}
\lab{b.1.Sigma}
\eea

The product of vertices $\Ga\Lb\la\,\vert\,\si,\De\Rb\Ga\Lb\la\prm\,\vert\,\si,\De\Rb$ that appears in \eq{b.1.Sigma},
contains two poles in the $\si$-plane in the following regions (for details see Eqs. (\ref{2.om1nu} - \ref{2.om3nu}) and \eq{lanu}):

\bea
&&\si\he\la\hspace{1.6cm}\Leftrightarrow\hspace{1cm} i/2-\nu\he\nu_1+\nu_2\lab{b.1.ourregion}\\
\nn\\
&&\si\he i-\la\hspace{1cm}\Leftrightarrow\hspace{1cm} i/2 + \nu\he\nu_1+\nu_2\lab{b.2.ourregion}
\eea

 We need to check that the vertices $\Om$ and $\La$ converge as $\nu\to \infty$,
 for both of the regions of \eq{b.1.ourregion} and \eq{b.2.ourregion}. 
From  \eq{lanu} its clear that  $\La$  falls down for pure imaginary $\nu = i \kappa$, as $\exp\Lb- 2 \,\kappa \ln \kappa \Rb$. 
Actually, we will see below that we need to integrate
  the $\nu$-image of the amplitude to calculate the scattering amplitude in  coordinate space, in the following way:

\beq \label{NUTOZETA}
A\Lb \zeta \Rb\,\,=\,\,\int \frac{d \nu}{2 \pi}\,e^{ i \nu \ln \zeta}\,A\Lb \nu\Rb
\eeq

One can see, that independent from the sign of $\ln \zeta$, we have to close the $\nu$-contour of integration for $\La$ over the upper half-plane.
Since the pole of \eq{b.1.ourregion} lies in the upper half-plane, then the $\nu$-integral for the part of the vertex proportional to $\La$,  
is taken using the contour which is closed around the pole given by \eq{b.1.ourregion}. \\

The situation with the part of the vertex proportional to $\Om$ is quite different.
One can see that at large and pure imaginary $\nu$, then $\Om$ falls down only as a power of $\nu$. Therefore, the convergence of the integral in \eq{NUTOZETA} depends on the sign of $\ln \zeta$.
We choose $\zeta > 1$ . For this choice of $\zeta$ we can close the $\nu$-integration contour  over the upper half-plane for both $\La$ and $\Om$.\\

With this in mind,
$\la\to 0$ from above the real axis, is our region of interest for calculating the contribution to the vertex proportional to $\La$.
Whereas $ \la\to 0$ from below the real axis, is the relevant region for calculating the part of the vertex proportional to $\Om$. The dominant high energy
behaviour $\exp\Lb 2\om_0\,Y\Rb$ stems from the region $\nu_1,\nu_2\to 0$, (or in other words $\si\to 0$).
Thus since $\la\to 0^+$ is our region of interest, then \eq{b.1.ourregion} is the relevant region, whereas \eq{b.2.ourregion} does not
yield $\si\to 0$ as $\la\to 0^+$. 
In light of this observation,
 it is useful to extract the pole explicitly by making the following definition:

\bea&&
\Ga\Lb\la\,\vert\,\si,\De\Rb\he\ml{\f{\tilde{\Ga}\Lb\la\,\vert\,\si,\De\Rb}{\la-\si}}\lab{b.3.Gammatilde}\eea

In this approach,  $\tilde{\Ga}\Lb\la\,\vert\,\si,\De\Rb$  is finite in the region $\si\to \la$.  
Hence plugging \eq{b.3.Gammatilde} into \eq{b.1.Sigma} leads to the following formula:

\bea&&
m\Lb\om,\la,\la\prm\Rb\he\Lint^\infty_{-\infty}d\si\Lint^\infty_{-\infty}d\De\,\, 
\,\f{g\Lb\ml{\f{\si +\De}{2}}\Rb g\Lb\ml{\f{\si -\De}{2}}\Rb}{\Lb\,\om-\om\Lb\ml{\f{\si+\De}{2}}\Rb-\om\Lb\ml{\f{\si -\De}{2}}\Rb\,\Rb}\,\,
\f{\tilde{\Ga}\bl\la\,\begin{vmatrix}\,\si,\De\br\tilde{\Ga}\bl\la\prm\,\end{vmatrix}\,\si,\De\br}
{\bl\la-\si\br\bl\la\prm-\si\br}\hspace{1cm}
\lab{b.2.Sigma}\\
\nn\\
&&\He -2\pi i \Lint^\infty_{-\infty}d\De\,\, \left\{
\,\f{g\Lb\ml{\f{\si +\De}{2}}\Rb g\Lb\ml{\f{\si -\De}{2}}\Rb}{\Lb\,\om-\om\Lb\ml{\f{\si +\De}{2}}\Rb-\om\Lb\ml{\f{\si -\De}{2}}\Rb\,\Rb}\,\,
\f{\tilde{\Ga}\bl\la\,\begin{vmatrix}\,\si,\De\br\tilde{\Ga}\bl\la\prm\,\end{vmatrix}\,\si,\De\br}
{\bl\la\prm-\la\br}\,\,\right\}_{\si\,=\,\la}
\hspace{1cm}
\lab{b.3.Sigma}\eea

where in the last step the $\si$-integral was solved by taking the residue of the pole at $\si\he\la$.
Switching back to $\Lb\nu,\nu_1,\nu_2\Rb$ variables, then \eq{b.3.Sigma} reads:

\bea
&&m\Lb\om,\nu,\nu\prm\Rb\he -2\pi i \Lint^\infty_{-\infty}d\nu_2  \left\{
\f{g\Lb\nu_1\Rb g\Lb\nu_2\Rb}{\bl \om-\om\Lb\nu_1\Rb-\om\Lb\nu_2\Rb \br }\,\,
\f{\tilde{\Ga}\bl\nu\,\begin{vmatrix}\,\nu_1,\nu_2\br\tilde{\Ga}\bl\nu\prm\,\end{vmatrix}\,\nu_1,\nu_2\br}
{\bl\nu\prm-\nu\br}\right\}_{i/2-\nu\, =\,\nu_1+\nu_2}
\lab{b.3a.Sigma}\eea

Interestingly, the Pomeron self-mass $m\Lb\om,\nu,\nu\prm\Rb$ contains a simple (first order) pole  in the region
$\nu=\nu\prm$. Hence using the result of \eq{b.3a.Sigma}:

\bea
&&
\Lint^\infty_{-\infty}d\nu\prm m\Lb\om,\nu,\nu\prm\Rb G_0\Lb\om,\nu\prm\Rb=
  4\pi^2\Lint^\infty_{-\infty} d\nu_2 \left\{ 
  \f{g\Lb\nu_1\Rb g\Lb\nu_2\Rb\,\tilde{\Ga}^2\bl\nu\,\vert\,\nu_1,\nu_2\br}{\bl \om-\om\Lb\nu_1\Rb-\om\Lb\nu_2\Rb \br }\,G_0\Lb\om,\nu\Rb\right\}_{\!i/2-\nu\,=\,\nu_1+\nu_2}\, \hspace{0.5cm}
\lab{b.3c.Sigma}
\eea

where the right hand side of \eq{b.3c.Sigma} was derived by taking the residue of the pole at $\nu\,=\,\nu\prm$, after integrating over $\nu\prm$.
The right hand side of \eq{b.3c.Sigma} can be re-written in the following equivalent form:

\bea&&\Lint^\infty_{-\infty}d\nu\prm m\Lb\om,\nu,\nu\prm\Rb G_0\Lb\om,\nu\prm\Rb\he\Sigma\Lb\om,\nu\Rb G_0\Lb\om,\nu\Rb\lab{useful}\\
\nn\\
\nn\\
&&
\Sigma\Lb\om,\nu\Rb\he
\Lint^\infty_{-\infty}d\nu_1\,\Lint^\infty_{-\infty}d\nu_2\,\,\ml{\f{g\Lb\nu_1\Rb\,g\Lb\nu_2\Rb}
   {\bl\,\om-\om\Lb\nu_1\Rb-\om\Lb\nu_2\Rb\,\br}} \,\f
{\tilde{\Ga}^2\biggl(\nu\,\vert\,\nu_1,\nu_2\biggr)}
{\Lb \ml{\f{i}{2}} -\nu-\nu_1-\nu_2\Rb\,}
\,\,
\lab{1.definitionSigma}\eea

where the $\nu_1$ integral in \eq{1.definitionSigma} is solved  by taking the residue of 
 the pole at $i/2 - \nu-\nu_1-\nu_2=0$.
Thanks to the simplification of \eq{useful}, then \eq{1loopomega} can be re-cast as:
 
\bea
&&
A_{(1)}\Lb\om,\nu\Rb\he G_0\Lb\om,\nu\Rb \Sigma\Lb\om,\nu\Rb\,\,G_0\Lb\om,\nu\Rb \lab{1.1loopomega}\eea

In order to pass to $Y$ representation, use the following inverse Mellin transform:

\bea
&&A_{(1)}\Lb Y,\nu\Rb\he\f{1}{2\pi i}\Lint^{a+i\infty}_{a-i\infty}d\om\,\,\, e^{\om\,Y}A_{(1)}\Lb\om,\nu\Rb\lab{0.1loopY}\eea

which after inserting Eq. (\ref{1.1loopomega})  yields:

\bea
&&A_{(1)}\Lb Y,\nu\Rb\he
\,\,g^2\Lb\nu\Rb\Lint^\infty_{-\infty}d\nu_1\,\Lint^\infty_{-\infty}d\nu_2\,\f{g\Lb\nu_1\Rb g\Lb\nu_2\Rb}{\bl \om-\om\Lb\nu_1\Rb-\om\Lb\nu_2\Rb\br}
\f
{\tilde{\Ga}^2\biggl(\nu\,\vert\,\nu_1,\nu_2\biggr)}
{\Lb \ml{\f{i}{2}} -\nu-\nu_1-\nu_2\Rb\,}
\hspace{1cm}\lab{1loopY}\\
\nn\\
&&\times\,\ml{\f{e^{\om\Lb\nu\Rb\,Y}}{\om\Lb\nu\Rb-\om\Lb\nu_1\Rb-\om\Lb\nu_2\Rb}}\,\Lb\,\ml{\f{e^{\left\{\,\om\Lb\nu_1\Rb+\om\Lb\nu_2\Rb-\om\Lb\nu\Rb\,\right\}\,Y}\,-1\,}{\om\Lb\nu\Rb-\om\Lb\nu_1\Rb-\om\Lb\nu_2\Rb}}
\,+Y\,\Rb
\nn\eea

\subsection{Two loops}

\DOUBLEFIGURE[t]{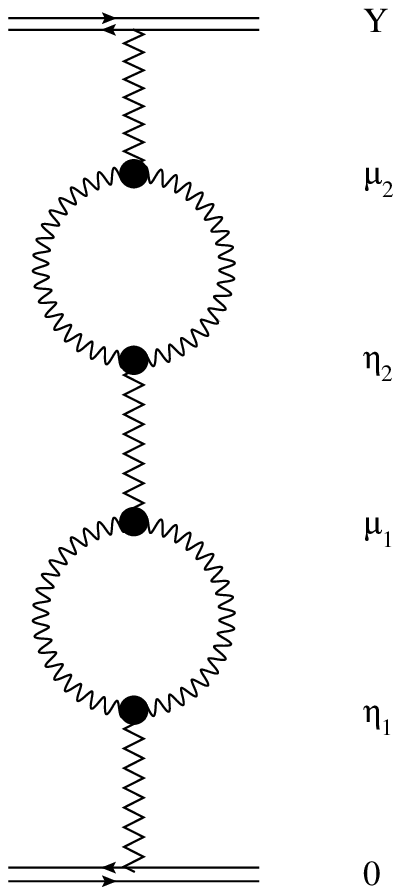,width=45mm,height=80mm}{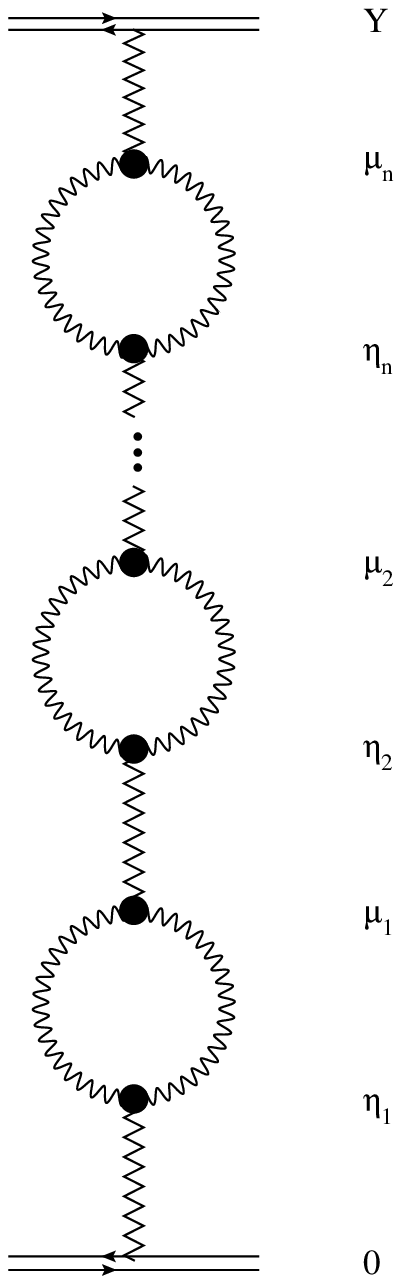,width=35mm,height=80mm}
{The diagram with two Pomeron loops.
\label{f2loop}}{The diagram with $n$ Pomeron loops in succession. \label{fnloop} }

The second correction to the basic diagram is the diagram with two loops
shown below in \fig{f2loop}. 
The amplitude for the 2-loop diagram takes the following form in $\om$ representation:

\bea&&
A_{(2)}\Lb\om,\nu\Rb\he G_0\Lb\om,\nu\Rb \Lint^\infty_{-\infty}d\nu\prm\Lint^\infty_{-\infty}d\nu\dpr 
\,\,m\Lb\om,\nu,\nu\prm\Rb \,\,G_0\Lb\om,\nu\prm\Rb\,\, m\Lb\om,\nu\prm,\nu\dpr\Rb\,\,
G_0\Lb\om, \nu\dpr\Rb
\hspace{1cm}\lab{0.2loopomega}\eea

Thanks to the useful result of \eq{useful}, 
the amplitude for the 2-loop diagram of \eq{0.2loopomega} 
simplifies to the following formula:

\bea&&
A_{(2)}\Lb\om,\nu\Rb\he G_0\Lb\om,\nu\Rb\,\,\bl\,\,G_0\Lb\om,\nu\Rb\,\,\Sigma\Lb\om,\nu\Rb\,\,\br^2\lab{2loopomega}\eea

where $\Sigma\Lb\om,\nu\Rb$ is given by \eq{1.definitionSigma}.
In order to pass to $Y$ representation, use the following inverse Mellin transform:

\bea&&
A_{(2)}\Lb Y,\nu\Rb\he\f{1}{2\pi i}\Lint^{a+i\infty}_{a-i\infty}d\om\,\,\, e^{\om\,Y}A_{(2)}\Lb\om,\nu\Rb\,\,\,\lab{2loopY}\\
&&\He
g^3\Lb\nu\Rb \prod^2_{k=1}\,\,\,\Lint^\infty_{-\infty}d\nu^k_1\,\Lint^\infty_{-\infty}d\nu^k_2\,\,\,g\Lb\nu^k_1\Rb\,g\Lb\nu^k_2\Rb
\f
{\tilde{\Ga}^2\biggl(\nu\,\vert\,\nu^k_1,\nu^k_2\biggr)}
{\Lb \ml{\f{i}{2}} -\nu-\nu^k_1-\nu^k_2\Rb\,}
\lab{toowide}\\
&&\times\hspace{0.3cm}\left\{\hspace{0.3cm} \ml{\f{e^{\om_1\,Y}}{\om-\om_1}}+\ml{ \f{e^{\om_2\,Y}}{\om-\om_2}} + 
e^{\om\Lb\nu\Rb\,Y}\Lb\ml{\f{1}{\Lb\om-\om_1\Rb\Lb\om-\om_2\Rb}}\Rb\,\f{1}{2!}\right.\nn\\
&&\left.\hspace{0.7cm}+\hspace{0.3cm}
 e^{\om\Lb\nu\Rb\,Y}\f{d}{d\om}\Lb\ml{\f{1}{\Lb\om-\om_1\Rb\Lb\om-\om_2\Rb}}\Rb \,Y+e^{\om\Lb\nu\Rb\,Y}\f{d^2}{d\om^2}\Lb\ml{\f{1}{\Lb\om-\om_1\Rb\Lb\om-\om_2\Rb}}\Rb\f{Y^2}{2!}\hspace{0.3cm}\right\}_{\omega \,=\, \omega(\nu)}\nn
\eea
\bea
&&\om_k\he\om\Lb\nu_1^k\Rb+\om\Lb\nu_2^k\Rb\lab{omk}\eea

\begin{boldmath}
\subsection{$n$ loops}\lab{snloops}
\end{boldmath}

Extending this approach to the diagram with $n$ loops in succession
shown in \fig{fnloop}, leads to the following amplitude in $\om$ representation:

\bea&&
A_{(n)}\Lb\om,\nu\Rb\he 
G_0\Lb\om,\nu\Rb\,\,\bl\,\,G_0\Lb\om,\nu\Rb\,\Sigma\Lb\om,\nu\Rb\,\,\br^n\lab{1.nloopomega}\eea

The sum over the class of diagrams shown in \fig{fnloop} with an alternating minus sign,
for all $n \in \Lb 0,\infty\Rb$ i.e. from $n=0$ loops up to infinity gives the Green function of the dressed Pomeron,  labeled $G_2\Lb\om,\nu\Rb$.
In this notation:

\bea&&G_2\Lb\om,\nu\Rb\he
\sum^\infty_{n=0}\Lb -1\Rb^nA_{(n)}\Lb \om,\nu\Rb\,\,\,
=\hspace{0.3cm}
\f{G_0\Lb\om,\nu\Rb}{\bl \hspace{0.2cm}1\,\,+\,\,G_0\Lb\om,\nu\Rb\,\Sigma\Lb\om,\nu\Rb\,\,\hspace{0.2cm}\br} \lab{1.G2} \eea

From \eq{1.G2}, the renormalized propagator $G_2\Lb\om,\nu\Rb$ can be expressed in terms of the bare Pomeron propagator $G_0\Lb\om,\nu\Rb$ as:

\beq
G_2^{-1}\Lb\om,\nu\Rb\he G_0^{-1}\Lb\om,\nu\Rb\hspace{0.3cm}+\hspace{0.3cm}\Sigma\Lb\om,\nu\Rb\lab{G2G0}\eeq

\begin{boldmath}
\subsection{Calculation of $\Sigma\Lb\om,\nu\Rb$}\lab{sSigma}
\end{boldmath}

In this section, we calculate explicitly the formula for the Pomeron loop  $\Sigma\Lb\om,\nu\Rb$, defined above in \eq{1.definitionSigma}\footnote{Throughout  the paper the notation $\lim_{x \to a} f(x)$ is used to denote the asymptotic behaviour of f(x) at $x \to a$. It is not meant in the conventional way as used in analysis} .
In ($\la,\si,\De$)
 notation introduced in  Eqs. (\ref{depm}):

\bea&&
\Sigma\Lb\om,\la\Rb\he\Lint^\infty_{-\infty}d\si\Lint^\infty_{-\infty}d\De\,\, 
\,\f{g\Lb\ml{\f{\si+\De}{2}}\Rb g\Lb\ml{\f{\si -\De}{2}}\Rb}{\Lb\,\,\,\om-\om\Lb\ml{\f{\si+\De}{2}}\Rb-\om\Lb\ml{\f{\si-\De}{2}}\Rb\,\,\,\Rb}\,\,
\f{\tilde{\Ga}^2\bl\la\,\vert\,\si,\De\br}
{\bl\la-\si\br}\lab{2.Sigma}\\
\nn\\
&&\He -2\pi i\hspace{0.3cm} \ml{\ml{\lim_{\si\to \la}}}\hspace{0.3cm}\Lint^\infty_{-\infty}d\De\,\,
\,\f{g\Lb\ml{\f{\la +\De}{2}}\Rb g\Lb\ml{\f{\la -\De}{2}}\Rb}{\Lb\,\,\,\om-\om\Lb\ml{\f{\la +\De}{2}}\Rb-\om\Lb\ml{\f{\la -\De}{2}}\Rb\,\,\,\Rb}\,\,
\tilde{\Ga}^2\bl\la\,\vert\,\si,\De\br\lab{3.Sigma}\eea

where in the last step, the integration over $\si$ was solved by taking the residue of the pole at
$\si\he\la$.
On the RHS of  \eq{1.G2},
 there are singularities at $i/2+\nu\to 0$ and $i/2-\nu\to 0$ that stem from $G_0\Lb\om,\nu\Rb$ in the numerator (see definition of \eq{gnu}).
However as  discussed above,  only the region $i/2-\nu\he\la\to 0$ is relevant (see \eq{b.1.ourregion} and the surrounding discussion). 
In light of this, the largest contribution to the propagator of the dressed Pomeron $G_2\Lb\om,\nu\Rb$ stems from the region  $i/2-\nu\he\la\to 0$. 
 Hence we need to know the asymptote of $\Sigma\Lb\om,\nu\Rb$ in this region
which can be found from \eq{3.Sigma}:


\bea&&\ml{\ml{\lim_{\la\to 0}}}\hspace{0.3cm}
\Sigma\Lb\om,\la\Rb\he -2\pi i\hspace{0.3cm} \ml{\ml{\lim_{\substack{\si\to\la\\\la\to 0}}}}\hspace{0.3cm}
\Lint^\infty_{-\infty}d\De\,\, 
\,\f{g\Lb\ml{\f{\la +\De}{2}}\Rb g\Lb\ml{\f{\la -\De}{2}}\Rb\,\,}
{\Lb\,\,\,\om-\om\Lb\ml{\f{\la +\De}{2}}\Rb-\om\Lb\ml{\f{\la -\De}{2}}\Rb\,\,\,\Rb}\,\,\,
\tilde{\Ga}^2\bl\la\,\vert\,\si,\De\br\,\hspace{1cm} \lab{5.Sigma}\eea
Throughout  the paper the notation $\lim_{x \to a} f(x)$ is used to denote the asymptotic behaviour of f(x) at $x \to a$. It is not meant in the conventional way as used in analysis}
 It is worthwhile mentioning
here the two contributions to the vertex $\tilde{\Ga}\Lb\la\,\vert\,\si,\De\Rb$ in this region.
Recall that from the definition of \eq{tpv} and \eq{b.3.Gammatilde} that:

\bea
\tilde{\Ga}\Lb\,\la\,\vert\,\si,\De\,\Rb\he \f{16\bas^2}{N_c}\Lb\la-\si\Rb\Lb\,\Ga_{\mbox{\footnotesize{planar}}}\Lb\la\,\vert\,\si,\De\Rb-\f{2\pi}{N_c^2}\,\,
\Ga_{\mbox{\footnotesize{nonplanar}}}\Lb\la\,\vert\,\si,\De\Rb\,\Rb\lab{1.pnp}\eea

From Eqs. (\ref{planartend}) and (\ref{nonplanartend}) the asymptotes $
\Lb\la-\si\Rb\Ga_{\mbox{\footnotesize{planar}}}\Lb\la\,\vert\,\si,\De\Rb$ and 
$
\Lb\la-\si\Rb\Ga_{\mbox{\footnotesize{nonplanar}}}\Lb\la\,\vert\,\si,\De\Rb$
in the narrow region where $\la\to \si$ and $\si\to 0$, are the same up to a numerical coefficient. 
With this in mind, thanks to the suppression factor of
$2\pi/N_c^2$ in front of $
\Ga_{\mbox{\footnotesize{nonplanar}}}\Lb\la\,\vert\,\si,\De\Rb$ in \eq{1.pnp},
the contribution of the non-planar diagram to the vertex is parametrically smaller by a factor of $2\pi/N_c^2$ than the planar
diagram, in our region of interest.
Nevertheless for the sake of completeness, the contribution of both diagrams to the vertex are included
in the calculation of $\Sigma\Lb\om,\la\Rb$.
Now inserting the asymptote of \eq{2.asymptotep} into \eq{5.Sigma} leads to the following result:

\bea&&\ml{\ml{\lim_{\la\to 0}}}\hspace{0.2cm}
\Sigma\Lb\om,\la\Rb\he -2\pi ia^2\,\ml{\ml{\lim_{\substack{\la\to 0}}}}\hspace{0.2cm}
\Lint^\infty_{-\infty}d\De
\,\f{g\Lb\ml{\f{\la +\De}{2}}\Rb g\Lb\ml{\f{\la -\De}{2}}\Rb}
{\Lb\,\,\om-\om\Lb\ml{\f{\la +\De}{2}}\Rb-\om\Lb\ml{\f{\la - \De}{2}}\Rb\,\,\Rb}\,\,\Lb\,\,\f{\la^2}{  \bl\la+\De\br\bl\la-\De\br}\,\,\Rb^2\hspace{1cm}
\lab{6.sigma}
\eea

where the numerical coefficient $a$ is given in \eq{definitionap}.  
Assuming that the typical value of $\De$ is small, then\footnote{Throughout  the paper the notation $\lim_{x \to a} f(x)$ is used to denote the asymptotic behaviour of f(x) at $x \to a$. It is not meant in the conventional way as used in analysis} \eq{6.sigma} can be re-cast as follows:

\bea&&\ml{\ml{\lim_{\la\to 0}}}\hspace{0.2cm}
\Sigma\Lb\om,\la\Rb\,\,=\,\, -2\pi ia^2\ml{\ml{\lim_{\substack{\la\to 0}}}}\hspace{0.2cm}
\Lint^\infty_{-\infty}\!\!d\De
\,\f{g\Lb\ml{\f{\la +\De}{2}}\Rb g\Lb\ml{\f{\la -\De}{2}}\Rb}
{\om\dpr\Lb \ml{\f{\la}{2}}\Rb\bl\De-\De_+\br\bl\De-\De_-\br}\,\,\Lb\,\,\f{\la^2}{\bl\la+\De\br\bl\la-\De\br}\,\,\Rb^2\hspace{0.5cm}
\lab{7.Sigma}\\
&&
\De_\pm\he\ml{\f{\pm\bl \om-2\om\Lb\ml{\f{\la}{2}}\Rb\br^{\lh}}{\om^{\,\prime\prime}\Lb\ml{\f{\la}{2}}\Rb}}
\lab{Depm}\eea

After closing the $\De$ integration contour over the upper half -plane that encloses the pole at $\De=\De_+$, and taking the residue in this region, then \eq{7.Sigma}
simplifies to:

\bea&&\ml{\ml{\lim_{\la\to 0}}}\hspace{0.2cm}
\Sigma\Lb\om,\la\Rb\he 4\pi^2 a^2\,\ml{\ml{\lim_{\substack{\la\to 0}}}}\hspace{0.2cm}
\,\f{g\Lb\ml{\f{\la +\De_+}{2}}\Rb g\Lb\ml{\f{\la -\De_+}{2}}\Rb}
{\om\dpr\Lb \ml{\f{\la}{2}}\Rb\bl\De_+-\De_-\br}\,
\,\,\,\Lb\,\,\f{\la^2}{\bl\la +\De_+\br\bl\la-\De_+\br}\,\,\Rb^2
\nn\\
\nn\\\nn\\
&&\He 2\pi^2 a^2\,\ml{\ml{\lim_{\substack{\la\to 0}}}}\hspace{0.2cm}
\,\f{g\Lb\ml{\f{\la +\De_+}{2}}\Rb g\Lb\ml{\f{\la -\De_+}{2}}\Rb}
{\Lb\om-2\om\Lb\ml{\f{\la}{2}}\Rb\Rb^{\lh}}\,
\,\,\,\Lb\,\,\f{\la^2}{\bl\la +\De_+\br\bl\la-\De_+\br}\,\,\Rb^2\lab{9.Sigma}
\eea

Assuming that $\De_+$ is small (i.e. in the region $\om\to 2\om\Lb\la/2\Rb$), then as $\la\to 0$:

\bea
\ml{\ml{\lim_{\la\to 0}}}\hspace{0.3cm}
g\Lb\ml{\f{\la +\De_+}{2}}\Rb\to4\Lb\la +\De_+\Rb^2\hspace{3cm}\Lb\,\De_+\ll 1\,\Rb\lab{gdl}\eea

where the definition of \eq{gnu} was used, with a similar result for $g\Lb\,\,\Lb\la-\De_+\Rb/2\,\,\Rb$. Hence
in the narrow region that $\om\to 2\om\Lb\la/2\Rb$ and $\la\to 0$,  then \eq{9.Sigma} reduces to:

\bea&&\ml{\ml{\lim_{\substack{\la\to 0\\\om\to 2\om\Lb\la/2\Rb}}}}\hspace{0.3cm}
\Sigma\Lb\om,\la\Rb\he
 32\pi^2 a ^2\,\,\,\ml{\ml{\lim_{\substack{\la\to 0\\\om\to2\om\Lb\la/2\Rb}}}}\hspace{0.3cm}
\,\f{\la^4}
{\sqrt{\om-2\om\Lb\ml{\f{\la}{2}}\Rb}}\,\,\,
\hspace{1cm}\lab{10.Sigma}\eea

for small $\De_+$,  as $\om\to2\om\Lb\la/2\Rb$. 

\subsection{Green function of the dressed Pomeron}
The Green function  of the dressed Pomeron can be calculated using \eq{G2G0},
which can be reduced to the following expression

\beq \lab{G2-1}
G_2^{-1}\Lb\om,\nu\Rb\,=\,  \frac{1}{g(\nu)} \Lb  \om \,\,-\,\,\omega\Lb \nu\Rb \,+\, g\Lb \nu\Rb
\Sigma\Lb\om,\nu\Rb \Rb \,\,\,\xrightarrow{ \lambda \to 0}\,\,-\,4\lambda^2 
\Lb \om \,\,-\,\,\omega\Lb \nu \Rb \,-\, \Lb 1/4\lambda^2\Rb
\Sigma\Lb\om,\lambda\Rb\Rb\eeq 

The singularities of the Green function stems from the following equation (substituting \eq{10.Sigma}):
\footnote{Throughout  the paper the notation $\lim_{x \to a} f(x)$ is used to denote the asymptotic behaviour of f(x) at $x \to a$. It is not meant in the conventional way as used in analysis.
\bea \lab{GFSI}
 &&\om \,\,-\,\,\omega\Lb \nu \Rb \,-\, \Lb 1/4\lambda^2\Rb
\Sigma\Lb\om,\lambda\Rb\,\,\he 0\lab{1.GFSI}\\
\nn\\
\Rightarrow\hspace{0.3cm}&&
 \om \,\,-\,\,\omega\Lb \nu \Rb\,\,-\,\,
 8\pi^2 a ^2\,\,\,\ml{\ml{\lim_{\substack{\la\to 0\\\om\to2\om\Lb \la /2\Rb}}}}\hspace{0.3cm}
\,\f{\la^2}
{\sqrt{\om-2\om\Lb\ml{\f{\la}{2}}\Rb}}\he 0 \lab{2.GFSI}\eea

One can see that in the region where $\omega > 2 \omega\Lb\f{\lambda}{2}\Rb$, the correction to the pole at $\om = \om(\nu)$ is small, and can be neglected.
However when  $\omega \to 2 \omega \Lb\f{\lambda}{2}\Rb$ this correction becomes large, and 
 then the dominant contribution in this region
 is:

\bea &&\ml{\ml{\lim_{\substack{\la\to 0\\\om\to2\om\Lb \la /2\Rb}}}}\hspace{0.3cm}
G_2\Lb\om,\nu\Rb\he \ml{\ml{\lim_{\substack{\la\to 0\\\om\to2\om\Lb \la /2\Rb}}}}\hspace{0.3cm}\ml{\f{1}{\Sigma\Lb\om,\nu\Rb}}\,\lab{2.G2G0}\eea

whereby substituting Eqs. (\ref{9.Sigma}) and (\ref{gdl}):

\bea
 &&\ml{\ml{\lim_{\substack{\la\to 0\\\om\to2\om\Lb \la /2\Rb}}}}
G_2\Lb\om,\nu\Rb\,= \ml{\ml{\lim_{\substack{\la\to 0\\\om\to2\om\Lb \la /2\Rb}}}}\,\,\left\{
 \f{\sqrt{\om-2\om\Lb\ml{\f{\la}{2}}\Rb}}
{32\pi^2 a^2 \bl \la+\De_+\br^2\bl\la-\De_+\br^2}\,\,\f{1}{ \Lb \ml{\f{\la^2}{ \Lb\la+\De_+\Rb\Lb\la-\De_+\Rb}}\Rb^2}\right\}
\nn\\
\nn\\&&
\he
 \f{\sqrt{\om-2\om\Lb\ml{\f{\la}{2}}\Rb}}
{32\pi^2 a^2\la^4\, }\,\,\hspace{1cm}
\lab{3.G2G0}\eea

Since $\De_+$ is small in the limit that $\om\to2\om\Lb \la /2\Rb$ (see \eq{Depm}) then \eq{3.G2G0} simplifies to the following
asymptotic formula:

\bea
 &&\ml{\ml{\lim_{\substack{\la\to 0\\\om\to2\om\Lb \la /2\Rb}}}}
G_2\Lb\om,\nu\Rb\,
\he\ml{\ml{\lim_{\substack{\la\to 0\\\om\to2\om\Lb \la /2\Rb}}}}\hspace{0.7cm}
\f{1}
{32\pi^2 a^2\,\la^4}\,\,\sqrt{\om-2\om\Lb\ml{\f{\la}{2}}\Rb}\lab{4.G2G0}\eea

\eq{4.G2G0} is the dominant part of the propagator of the dressed Pomeron. 
The behaviour of the dressed Pomeron propagator with energy can be seen by transforming to $Y$ representation using the following inverse Mellin transform:

\bea&&
A^{\mbox{dressed}} \Lb Y,\nu \Rb\he\Lint^{a+i\infty}_{a-i\infty}\f{d\om}{2\pi i}\,e^{\om\,Y}G_2\Lb\om,\nu\Rb\lab{DRPOY}\eea

\eq{DRPOY} leads to the amplitude  of the exchange of one dressed Pomeron, that grows with energy according to the following 
behaviour: 

\beq \lab{DRPO}
A^{\mbox{dressed}} \Lb Y,\nu\Rb \,\,\,\,\,\,\, \propto  \,\,\,\,\,\, \frac{1}{Y^{3/2} }\,e^{2 \om\Lb\la/2\Rb\,Y}
\eeq
We believe that we have learned two lessons from this re-summation. The first one is that the enhanced diagrams change the asymptotic behaviour of the scattering amThroughout  the paper the notation $\lim_{x \to a} f(x)$ is used to denote the asymptotic behaviour of f(x) at $x \to a$. It is not meant in the conventional way as used in analysis}plitude.
 The second  is that they contribute in the rather narrow region $
\la \to 0$ and $\om \to 2 \omega(\la/2)$.

\begin{boldmath}
\section{High energy asymptotic behaviour of the scattering amplitude}
\end{boldmath}
\subsection{The Pomeron interaction vertices}

The goal of this section is to sum over all enhanced diagrams, 
using a method based on 
 the example of the previous section. 
The aim of our technique is to show that the 
more general diagrams for the Pomeron self-mass $\Sigma\Lb\om,\nu\Rb$ shown in \fig{sigmas}, are equivalent to the
diagram of \fig{Eqsigma} after replacing the Pomeron $1\to 2$ vertex with the $1\to n$ vertex.
From a field theory perspective, when one of the diagrams in \fig{sigmas} is cut, a factor of 
 $ 1/\Lb \omega - \sum_i \omega(\nu_i)\Rb$ is included in the expression for $\Sigma\Lb\om,\nu\Rb$,
where the sum is over all the Pomerons in the cut, with BFKL kernel $\om\Lb\nu_i\Rb$.
In this approach, each cut  brings an additional pole  in the $\omega$-plane.
 $\Sigma\Lb\om,\nu\Rb$ can be transformed to $Y$ representation by the inverse Mellin transform:

\bea
\Sigma\Lb Y,\nu\Rb\he\f{1}{2\pi i}\Lint^{a+i\infty}_{a-i\infty}d\om\, e^{\om Y}\,\Sigma\Lb\om,\nu\Rb\lab{invmellin}\eea

Using  \eq{invmellin}, the contour of the $\om$-integral
 can be closed over each pole that stems from  $ 1/\Lb \,\omega - \sum_i \omega(\nu_i)\,\Rb$.
 The residue from each pole
 will lead to the expression for  $\Sigma\Lb Y,\nu\Rb\,\,\propto\,\,\exp \Lb \sum_i \omega(\nu_i) Y\Rb$ (where $Y$ is the energy variable
 for dipole - dipole scattering.)  
The largest contribution to $\Sigma\Lb Y,\nu\Rb$ stems from the pole $1/\Lb\,\om-\sum^n_{i=1}\om\Lb\nu_i\Rb\,\Rb$ in the $\om$-plane,
where $n$ is the maximum number of Pomerons, that can be cut in the diagram.
The residue of this pole yields the contribution to  $\Sigma\Lb Y,\nu\Rb$ of the order:

\beq \label{MAXCONT}
\Sigma_n\Lb Y,\nu\Rb \,\,\,\propto\,\, V^n \exp\Lb \,\, \sum^n_{i=1} \omega(\nu_i)\,Y\,\,\Rb\,\,\,\approx\,\,\,a^n e^{n \omega_0 Y}
\eeq

where $\omega_0 =  4\bas\ln 2$ is the
leading order contribution to the intercept of the BFKL Pomeron, (see the expansion of \eq{OMNU0} and the surrounding discussion). 
Based on this observation,
 we propose  the following method
for calculating the Pomeron self-mass $\Sigma\Lb\om,\nu\Rb$, for the general
diagrams of \fig{sigmas}. Consider the diagram, where the maximum number of Pomerons in a cut is $n$.
Then $\Sigma_n\Lb\om,\nu\Rb$ is proportional to:

\bea
\Sigma_n\Lb\om,\nu\Rb\hspace{0.3cm}\propto\hspace{0.3cm}\,\Lb\,\f{1}{\om-\om\Lb\nu_1\Rb}\,\Rb\,\Lb\,\f{1}{\om-\om\Lb\nu_1\Rb-\om\Lb\nu_2\Rb}\,\Rb
\,\dots\,\Lb\,\f{1}{\om-\sum^n_{i=1}\om\Lb\nu_i\Rb}\,\Rb
\lab{cuts}
\eea

where each pole  $1/\Lb\, \om-\sum_i \om\Lb\nu_i\Rb\,\Rb$ stems from a different cut in the diagram. The term
 $1/\Lb\,\om-\sum^n_{i=1}\om\Lb\nu_i\Rb\,\Rb$ comes from the cut, that cuts the maximum number ($n$)  Pomerons.
To transform to $Y$ representation, \eq{cuts} should be substituted into \eq{invmellin}. In our approach,
we close the $\om$-contour around the pole $1/\Lb\,\om-\sum^n_{i=1}\om\Lb\nu_i\Rb\,\Rb$ (the maximum Pomeron cut).
Then the solution is equal to the residue of this pole, i.e. we replace $\om\he\sum^n_{i=1}\om\Lb\nu_i\Rb$
everywhere in the integrand.
The remaining poles are absorbed in the expression for the $1\to n$ Pomeron vertex, (which we will
derive below).
In this way the diagrams of \fig{sigmas} are equivalent to the diagram shown in \fig{Eqsigma}.
In this approach, $\Sigma_n$ can be calculated according to the following formula (which is shown
graphically in   \fig{Eqsigma}):

\beq \label{EQSI}
\ml{\Sigma}_n\Lb \omega, \nu\Rb\he \Lint\,
\ml{\prod}_i^n\, \frac{d \nu_i}{2 \pi i}\, \nu_i^2\,\Gamma\Lb  \nu\,\vert\, \{\nu_i\}\Rb\,\, \frac{1}{\om \,-\,\sum^n_{i=1} \,\om(\nu_i)}\,\,\Gamma\Lb  \nu\,\vert\, \{\nu_i\}\Rb
\eeq

where $\left\{\nu_i\right\}$ denotes $\nu_1,\nu_2,\dots\nu_n$.
This method of calculation, is directly related to the Mueller-Patel-Salam-Iancu approximation, 
for calculating the main contribution to the scattering amplitude due to the exchange of BFKL Pomerons \cite{MPSI}.
Strictly speaking, \eq{EQSI} is the $t$-channel unitarity constraint in $(\om, \nu)$-representation.

\DOUBLEFIGURE[t]{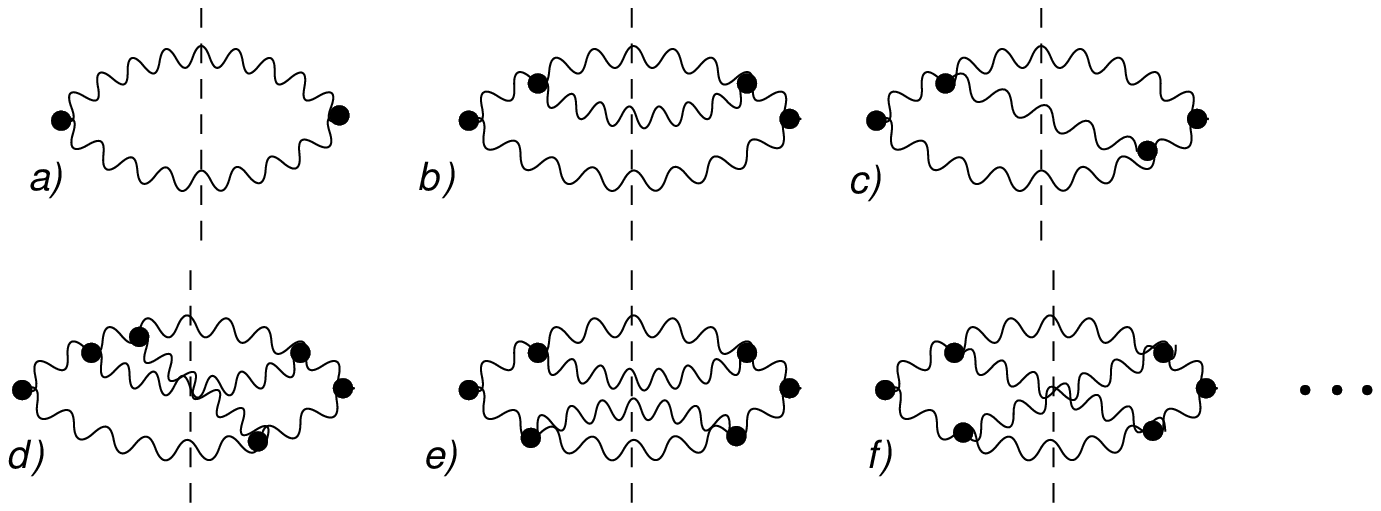,width=90mm,height=40mm}{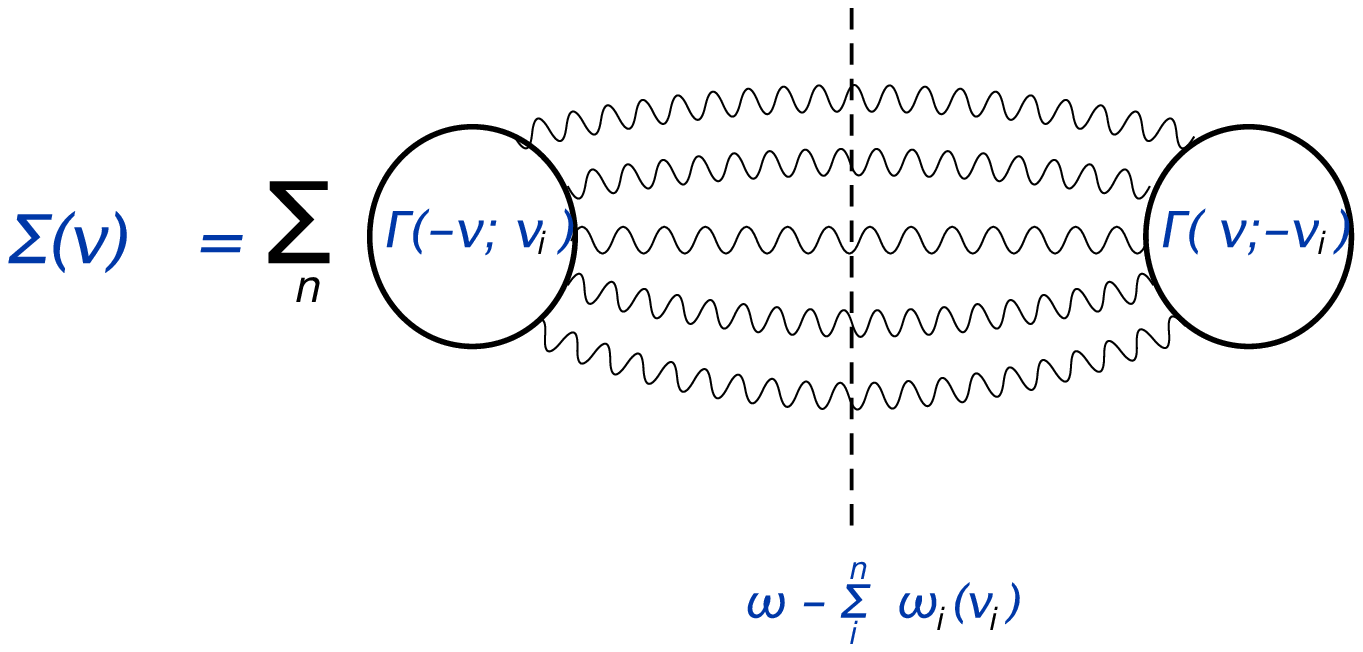,width=70mm,height=40mm}
{Examples of the diagrams for the Pomeron self-energy $\Sigma$. Dashed lines denote the cut with the maximal number of Pomerons. Wavy lines are used for BFKL Pomerons.
\label{sigmas}}{The graphical representation  for the formula of  \protect\eq{EQSI}, 
for the  Pomeron self-energy $\Sigma_n$. \label{Eqsigma} }

\FIGURE[h]{\begin{minipage}{160mm}
\centerline{\epsfig{file=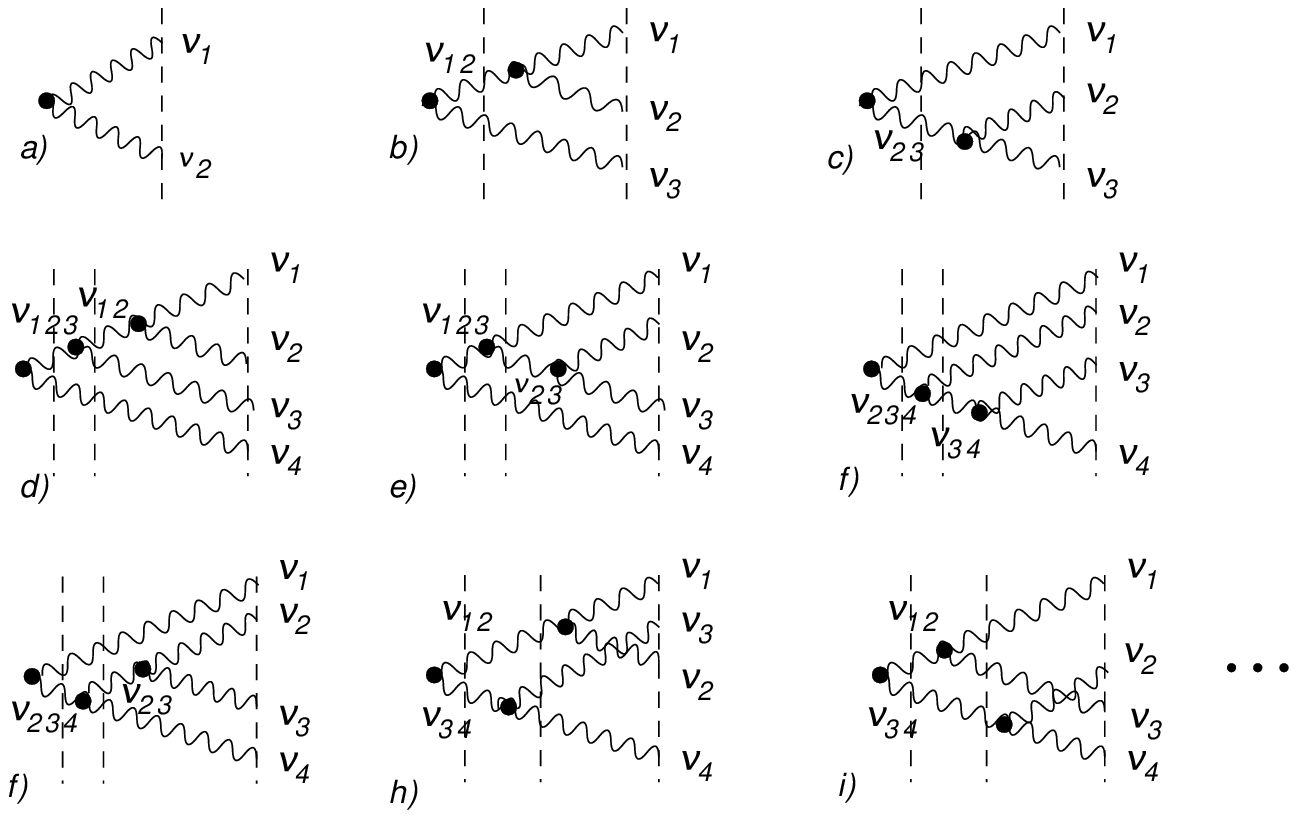,width=150mm}}\end{minipage}
\caption{The diagrams for the multi-Pomeron vertices. Fig. a) shows the simplest $1\to2$ vertex.
Fig. b) and Fig. c) show the $1\to3$ vertices, and Figs. d) - i) show the $1\to 4$  vertices. The dashed lines show the  cross sections with different energy ($\om$) propagators. Wavy lines describe the BFKL Pomerons.
} \label{vertices}}

The diagrams for the $1\to n$ Pomeron vertices are shown in \fig{vertices}.  The  simplest diagram for the $1 \to 2$ vertex
shown in \fig{vertices} a), was calculated in the previous section in detail.
It is useful to illustrate the main steps of this calculation,
 since this approach can be easily generalized to the calculation of the $1\to n$ vertex, for arbitrary $n$.
We draw attention to the formula for the $1\to 2$ vertex given in \eq{tpv}.
Recall that in this formula, $\nu$ is the  scaling dimension  of the Parent Pomeron, and
$\nu_1,\nu_2$ are the  scaling dimensions of the two daughter Pomerons, that are produced at the vertex.
The dominant contribution to the $1\to 2$ vertex
stems from the  singular region $i/2  - \nu\,=\, \nu_1 + \nu_2$. 
 Closing the integration contour around this pole, leads to the conservation relation:

\beq \lab{CONRE}
i/2 - \nu \,=\,\nu_1 + \nu_2\hspace{1cm}\mbox{or}\hspace{1cm}\lambda_{12}\,=\,\lambda_1 \,+\,\lambda_2
\eeq

where $\la_{12}\,=\,i/2-\nu$ and $\la_i\,=\,\nu_i\,\,\,(i=1,2)$.
The values of $\nu_1$ and $\nu_2$ are small, since this leads to the  dominant $Y$-dependence of the simple loop of \fig{f1loop}, proportional to
 $\exp\Lb 2 \om_0 Y\Rb$\footnote{The dominant contribution stems from small
$\nu_1$ and $\nu_2$, which are of the order $ \ln (x^2)/\sqrt{\om\dpr(0)\,Y}\,\,\ll\,\,1$, where $x^2$ is the variable, built from the size of the dipole 
(see ref.\cite{LMP} for the formula for $x^2$).}.  Hence from \eq{CONRE} we can conclude that $\nu \to i/2$, (or in other words $i/2-\nu =\lambda_{12} \to 0$) at the $1\to 2$ vertex.\\

Now we generalize to the notation  $\nu_i$ $(i=1,2,\dots, n)$, where $\nu_i$  denotes the  scaling dimension of the
$n$ daughter Pomerons, produced from the $1\to n$ vertex.  The values of the $\nu_i$'s are small, which leads to 
the  $Y$-dependence of the diagrams proportional to
 $\exp\Lb n \om_0 Y\Rb$.  At the simplest $1\to 2$ vertex (see \fig{vertices} a) ) where $ \nu_{12}\to\nu_1,\nu_2$,
the conservation law $i/2-\nu_{12}=\nu_1+\nu_2$ holds. Hence for $\nu_1,\nu_2$ small, then $\nu_{12}=i/2$.
At the $1\to 3$ vertex shown for example in \fig{vertices} b)   $\nu_{123}\to\nu_{12},\nu_3$, and at the $1\to 4$ vertex shown for example
in \fig{vertices} h), 
$\nu_{1234}\to\nu_{12},\nu_{34}$.In general this leads to the conservation rule at large $n$:

\bea 
\nu_{12\dots n}=i\Lb n-1\Rb /2+\sum^n_{i=1}\la_i\hspace{1cm}\la_i\ll 1\lab{nconre}
\eea

where the  scaling dimensions of the produced Pomerons, are denoted by $\la_i$, where $ \lambda_i \ll 1$. Here $n$ is the integer number which counts the 
number of Pomerons produced  in the tree decay, that started with one Pomeron.
Within the general $1 \to n$
vertex diagram, $\nu_{12\dots n}$ is close to $i \Lb n - 1\Rb /2$. 
Note that \fig{vertices} a) contains one vertex, \fig{vertices} b) contains 2 vertices and \fig{vertices} d) contains 3 vertices, such that
one can generalize to the $1\to n$ vertex that contains $n-1$ sets of $1\to 2$ vertices.
The expression for the $1\to 3$ vertex diagrams, as illustrated in  \fig{vertices} b) and \fig{vertices} c) can be written  as follows:

\bea \lab{V13}&&
\Gamma\Lb \nu_{123}\,\vert \,\nu_1,\nu_2,\nu_3\Rb\he\\
\nn\\
&&\He
 \left\{\,\,\,  \frac{ \Gamma\Lb \nu_{123}\,\vert\, \nu_{12},\nu_3\Rb\,g\Lb \nu_{12}\Rb\,\Gamma\Lb \nu_{12}\,\vert\, \nu_1,\nu_2\Rb
}{\om \,\,-\,\,\om(\nu_{12} )\,\,-\,\,\om(\nu_3)} \,\hspace{0.3cm}+\hspace{0.3cm}
 \frac{\Gamma\Lb \nu_{123}\,\vert \,\nu_{23},\nu_1\Rb\,g\Lb \nu_{23}\Rb\, \,\Gamma\Lb \nu_{23}\,\vert\, \nu_2,\nu_3\Rb 
}{\om \,\,-\,\,\om(\nu_{23} )\,\,-\,\,\om(\nu_1)}\right\}_{\om =\sum^3_{i=1}\om(\nu_i) }
\nn\eea

In \eq{V13}, $\om=\sum^3_{i=1}\om\Lb\nu_i\Rb$ because we are taking the residue of the $\om$-plane pole $1/\Lb\om-\sum^3_{i=1}\om\Lb\nu_i\Rb\Rb$,
which comes from the cut in the diagrams of \fig{vertices} (b) and \fig{vertices} (c), that cuts all 3 Pomerons. \eq{V13} can be
 be written as

\bea \label{V131}
&\Gamma\Lb \nu_{123}\,\vert\,\nu_1,\nu_2,\nu_3\Rb\,\,=\,\,\,
  \Gamma\Lb \nu_{123}\,\vert\, \nu_{12},\nu_3\Rb\,\hat{\Ga}\Lb \nu_{12}\,\vert \,\nu_1,\nu_2\Rb
\,\,\,+\,\,
\Gamma\Lb \nu_{123}\,\vert\, \nu_{23},\nu_1\Rb \,\hat{\Gamma}\Lb \nu_{23}\,\vert\, \nu_2,\nu_3\Rb
\eea

where the following definition was introduced:

\bea
&&\hat{\Ga}\Lb\nu_{12}\,\vert\,\nu_1,\nu_2\Rb\he\f{g\Lb\nu_{12}\Rb\Ga\Lb\nu_{12}\,\vert\,\nu_1,\nu_2\Rb}{\om\Lb\nu_1\Rb +\om\Lb\nu_2\Rb-\om\Lb\nu_{12}\Rb}\,\lab{1.V132}
\eea

\FIGURE[h]{\begin{minipage}{140mm}{
\centerline{\epsfig{file=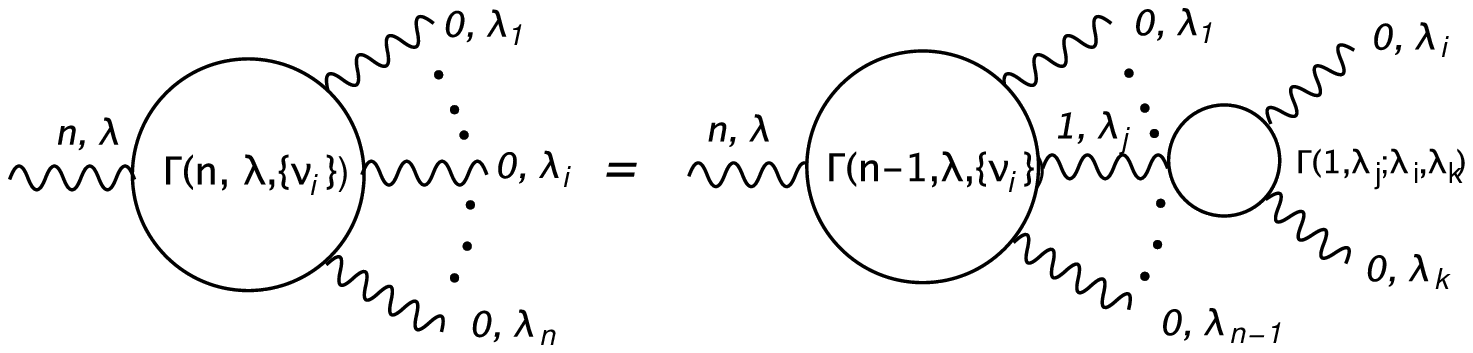,width=120mm}}}
\end{minipage}
\caption{ The graphic form of the equation for $\hat{\Ga}$, given in \eq{EQVN}
} \label{vertices1}}

with a similar definition for $\hat{\Ga}\Lb\nu_{23}\,\vert\,\nu_2,\nu_3\Rb$. The equation for the vertex  where 1 Pomeron $\to\,\,\, n$ Pomerons  is shown in \fig{vertices1}.
The equation for the vertex takes the following form: 

\bea   \label{EQVN}
&&\hat{\Ga}\Lb n, \lambda\,\vert\,\{\lambda_i=\nu_i\}\Rb \,\,\,=\,\,\,\sum^{n-1}_{j=1}\hat{\Ga}\Lb n-1,\lambda\,\vert\,\{\lambda_{i}\}, \lambda_j \Rb \, \hat{\Ga}\Lb 2,\lambda_j\,\vert\, \lambda_{i},\lambda_{k}\Rb
\eea

where 

\bea
\hat{\Ga}\Lb 2,\la_{i k}\,\vert\,\la_i,\la_k\Rb\he g \Lb\la_{i k}\Rb\f{\Ga\Lb\la_{i k}\,\vert\,\la_i,\la_k\,\Rb}{\om\Lb\la_{i k}\Rb -\om\Lb\la_i\Rb-\om\Lb\la_k\Rb}\lab{diff1}
\eea

This equation tells us, that the emission of one extra Pomeron (shown  by a zigzag line in \fig{fWI}),
can be reduced to the  emission of one extra Pomeron from the produced $(n-1)$ daughter Pomerons.
 Diagrams where the extra Pomeron is emitted elsewhere in the decay tree, cancel.
 Indeed, the vertex for one Pomeron emission  $\Gamma\Lb 2, \lambda|\nu_i, \nu_k\Rb$ can be re-written in the form

\beq \label{WI1}
\Gamma\Lb 2, \lambda_{i k}|\la_i, \la_k\Rb\,\,\,=\,\,\hat{\Gamma}\Lb 2, \lambda_{i k} |\la_i ,\la_k\Rb\,\Big\{\,\,G^{-1}\Lb \om, \Sigma^{(1)}\Rb\,\,-\,\,G^{-1}\Lb \om, \Sigma^{(2)}\Rb\,\,\Big\}
\eeq

 where, using the example of \fig{fWI} - B:

\bea
&& G^{-1}\Lb \om, \Sigma^{(1)}\Rb\he\om-\Sigma^{(1)}\he\om-\om\Lb\la_{012}\Rb-\om\Lb\la_3\Rb\lab{Sigma(1)}\label{SI1}\\
\nn\\
&& G^{-1}\Lb \om, \Sigma^{(2)}\Rb\he\om-\Sigma^{(2)}\he\om-\om\Lb\la_0\Rb-\om\Lb\la_{12}\Rb-\om\Lb\la_3\Rb\lab{Sigma(2)}\label{SI2}
\eea

Using this Ward identity we can show, that thanks to the cancellations of the diagrams, (as shown by the example of \fig{fWI}),
the emission of the extra Pomeron occurs  only from the produced $(n - 1)$ daughter Pomerons in the final state.
Diagrams where the extra Pomeron is produced from intermediate Pomerons in the decay tree, cancel. In \fig{fWI} we show the use of the Ward identity
 when calculating the $1 \to 4$ vertex, in terms of the  $1 \to 3$ vertex.
 The blob in \fig{fWI} is used to denote  the product of two vertices $\hat{\Ga}(1\to 2)\,\hat{\Ga}(1\to 2)$ without any $G(\om,\nu_{ik})$ between them. After summing all of the diagrams in \fig{fWI}-A,\fig{fWI}-B and 
\fig{fWI}-C, one can see that $\hat{\Gamma}(4,\lambda | \nu_1,\nu_2,\nu_3, \nu_4)\,\,=\,\,\sum_{i=1}^3\hat{ \Gamma}\Lb 3,\lambda | \lambda_{ik},\nu_l,\nu_j\Rb \hat{\Ga}\Lb 2 ,\lambda_{ik} | \nu_i, \nu_k \Rb
$ where $i \neq k \neq l \neq j$.

\FIGURE[h]{\begin{minipage}{140mm}{
\centerline{\epsfig{file=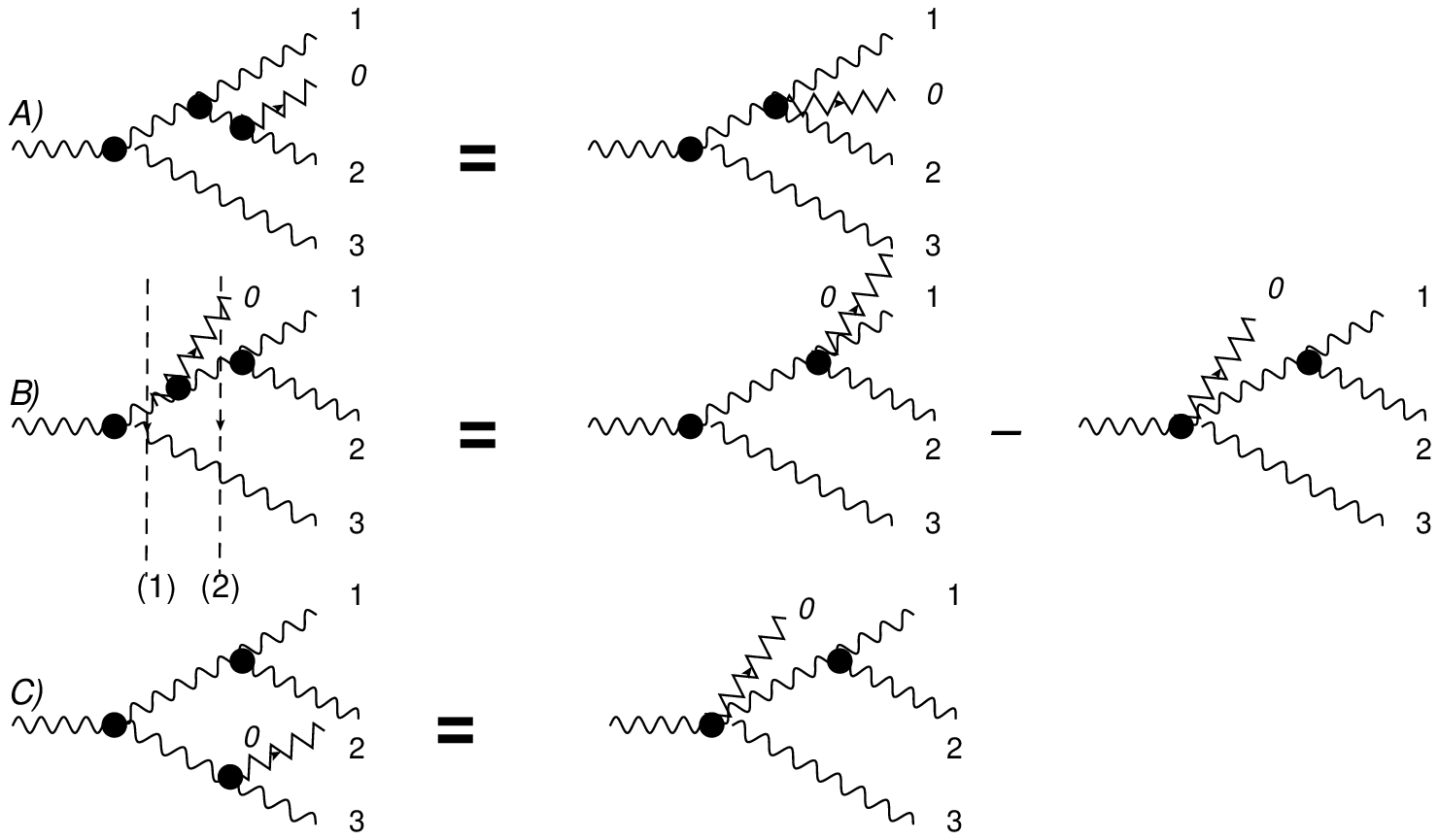,width=120mm}}}
\end{minipage}
\caption{ The illustration of the Ward identity of \protect{\eq{WI1}}. The extra emitted Pomeron is denoted by the zigzag line.
} \label{fWI}}

After this general outline of the calculation using \eq{WI1}, we  calculate the example of $\Ga\Lb 1 \to 4\Rb$ in more detail. 
Each diagram of \fig{fWI} can be written as a product  of three $\hat{\Ga}\Lb 2, \lambda_{i k}|\lambda_i, \lambda_k\Rb$ terms. Since 
$\hat{\Ga}\Lb 2, \lambda_{i k}|\lambda_i, \lambda_k\Rb\,\propto \la_{i k}/(\la_i \la_k)$ (see \eq{GA1} below), one can see that this product turns out to be
 the same for each of the diagrams in Fig.10. Having this in mind, and using \eq{WI1} we have for the diagram of \fig{fWI}-B
\bea \label{WI2}
\mbox{Fig. 10 - B}& \rightarrow &\hat{\Ga}\Lb 4, \la| \la_0,\la_1,\la_2,\la_3\Rb\,=\,\hat{\Ga}\Lb 2,\la | \la_{012},\la_3\Rb \hat{\Ga}\Lb \la_{012} | \la_{12}, \la_0\Rb \hat{\Ga}\Lb \la_{12} | \la_{1}, \la_2\Rb\nn\\
&\times& G\Lb \om, \Sigma^{(2)}\Rb \Big \{\,\,
G^{-1}\Lb \om, \Sigma^{(1)}\Rb\,\,-\,\,G^{-1}\Lb \om, \Sigma^{(2)}\Rb\,\,\Big\} G\Lb \om, \Sigma^{(1)}\Rb\,\,
G^{-1}\Lb \om, \Sigma^{(1)}\Rb\nn\\
&=& \prod^3_{i=1} \hat{\Ga}_i \Big\{ G\Lb \om, \Sigma^{(2)}\Rb\,G^{-1}\Lb \om, \Sigma^{(1)}\Rb\,-\,\,1\Big\} 
\eea

Here  $\prod^3_{i=1} \hat{\Ga}_i$ is used to denote  the product of corresponding $\hat{\Ga}$ terms, since it does not depend on the diagram.
The first term in \eq{WI2}  describes the emission from the produced Pomerons and has the same structure as the diagram of \fig{fWI}-A  at $\om = \sum^3_{i=0} \om\Lb \la_i\Rb$. Indeed,  $\hat{\Ga}$ for this  diagram has the form
\bea \label{WI3}
\mbox{Fig. 10 - A}& \rightarrow &\hat{\Ga}\Lb 4, \la| \la_0,\la_1,\la_2,\la_3\Rb\,=\,\hat{\Ga}\Lb 2,\la | \la_{012},\la_3\Rb \hat{\Ga}\Lb \la_{012} | \la_{02}, \la_1\Rb \hat{\Ga}\Lb \la_{02} | \la_{0}, \la_2\Rb\nn\\
&\times& G\Lb \om, \Sigma^{(2)}\Rb\,\Big( - \om\Lb \la_{0 1 2} \Rb + \om\Lb \la_{1}\Rb + \om\Lb\la_{02}\Rb\Big)
\frac{1}{ \om  - \om\Lb \la_{0 1 2} \Rb - \om\Lb \la_3\Rb} G^{-1}\Lb \om, \Sigma^{(1)}\Rb\nn\\
&\xrightarrow{ \om = \sum^3_{i=0} \om\Lb \la_i\Rb}& \prod^3_{i=1} \hat{\Ga}_i
  G\Lb \om, \Sigma^{(2)}\Rb\,G^{-1}\Lb \om, \Sigma^{(1)}\Rb
\eea
The difference between \eq{WI3} and the first term in \eq{WI2} is in the expression for $\Sigma^{(2)}$ which is equal to $\Sigma^{(2)}\,=\,\om\Lb \la_1\Rb + \om\Lb \la_{02}\Rb + \om\Lb \la_3\Rb$ instead of \eq{SI2}.

The second term in \eq{WI2} cancels with the diagram of \fig{fWI}-C.  This diagram is actually equal to sum of the two terms shown in \fig{vertices}-h and \fig{vertices}-i. This sum is equal to the following expression.
\bea \label{WI}
\mbox{Fig. 10 - c}& \rightarrow &\hat{\Ga}\Lb 4, \la| \la_0,\la_1,\la_2,\la_3\Rb\,=
 \prod^3_{i=1} \hat{\Ga}_i \frac{1}{ \om \,-\,\om\Lb\la_{12}\Rb - \om\Lb \la_{03}\Rb}\nn\\
 & \times&
 \left\{\Lb \om\Lb \la_{0} \Rb + \om\Lb \la_3\Rb - \om\Lb \la_{03}\Rb\Rb \frac{ \om\Lb \la_0\Rb + \om\Lb \la_3\Rb - \om\Lb \la_{03}\Rb}
{\om - \om\Lb \la_{03}\Rb - \om\Lb \la_1\Rb - \om\Lb \la_2\Rb}\right.\,\nn\\
  &+&\,\left.
 \Lb \om\Lb \la_{1} \Rb + \om\Lb \la_2\Rb - \om\Lb \la_{12}\Rb\Rb \frac{ \om\Lb \la_1\Rb + \om\Lb \la_2\Rb - \om\Lb \la_{12}\Rb}
{\om - \om\Lb \la_{12}\Rb - \om\Lb \la_0\Rb - \om\Lb \la_3\Rb}\right\} \nn\\
 &\xrightarrow{ \om = \sum^3_{i=0} \om\Lb \la_i\Rb}&\,\,\,\, \prod^3_{i=1} \hat{\Ga}_i
  \eea
Summarizing we see that  \eq{WI1} leads to the cancellation of the emission of the Pomeron from the internal lines in the diagram. Such cancellations are analogous to the cancellation in gauge theories since \eq{WI1} is similar to the Ward identity in these theories.

As discussed above, at the $1\to 2$ vertex, the conservation rule $i/2-\nu_j=\nu_i+\nu_k\,\, ( \la_{i k} = \la_i + \la_k)$ (see \eq{CONRE} and
the surrounding discussion).  We are interested in
the region where $\nu_i$ and $\nu_k$ are small, since this leads to the $Y$-dependence of the diagrams proportional to $\exp\Lb n\om_0 Y\Rb$, which is the dominant
contribution. Thus the relevant  region is $i/2-\nu_j\to 0$. 
Substituting for $\Ga\Lb\nu_j\,\vert\,\nu_i,\nu_k\,\Rb$ the formula of \eq{tpv},  then \eq{diff1} becomes:

\bea&&\hat{\Ga}\Lb 2,\nu_j\,\vert\,\nu_i,\nu_k\Rb\he\f{16\bas^2}{N_c}\,\,\,\,g\Lb\nu_j\Rb\,\,\,\,\f{\Lb\Ga_{\footnotesize{\mbox{planar}}}\Lb\nu_j\,\vert\,\nu_i,\nu_k\,\Rb - \ml{\f{2\pi}{N_c^2}} 
\Ga_{\footnotesize{\mbox{nonplanar}}}\Lb\nu_j\,\vert\,\nu_i,\nu_k\,\Rb\Rb}{\om(\nu_j)\,-\,\om(\nu_i)\,-\,\om(\nu_k)}
\lab{diff2}
\eea

Now to calculate the residue of $\hat{\Ga}\Lb 2,\nu_j\,\vert\,\nu_i,\nu_k\Rb$ in our region of interest, namely
$i/2-\nu_j-\nu_i-\nu_k\to 0$ when $\nu_i,\nu_k\to 0$, 
simply substitute  into \eq{diff2} the asymptotic formulae for $\Ga_{\footnotesize{\mbox{planar}}}$ and $\Ga_{\footnotesize{\mbox{nonplanar}}}$
derived in Eqs. (\ref{planartend}) and (\ref{nonplanartend}). Note that in this region, as $i/2-\nu_j\to\nu_i+\nu_k\to  0$ the denominator of \eq{diff2} tends to
$ \om(\nu_j)\,-\,\om(\nu_i)\,-\,\om(\nu_k)\to\bas/\Lb \h+i\nu_j\Rb$ (see definition of \eq{bfkl} ) and $g\Lb\nu_j\Rb\to -\Lb \h+i\nu_j\Rb^{-2}/4$ (see definition \eq{gnu}).
With this in mind, overall \eq{diff2} in this region reads:

\bea&&\lim_{\substack{i/2-\nu_j-\nu_i-\nu_k\to 0\\\nu_i,\nu_k\to 0}}\hspace{0.3cm}
\hat{\Ga}\Lb 2,\nu_j\,\vert\,\nu_i,\nu_k\Rb\he \lim_{\substack{i/2-\nu_j-\nu_i-\nu_k\to 0\\\nu_i,\nu_k\to 0}}\,b\,\f{\Lb \h+i\nu_j\Rb}{\nu_i\nu_k}\nn\\
\nn\\
&&\He\lim_{\nu_i,\nu_k\to 0}\hspace{0.3cm}
 b\,\f{\,\Lb \nu_i+\nu_k\Rb}{\nu_i\nu_k}\ \lab{GA1}\\
\nn\\
\nn\\
&&b\he \f{-a}{4\bas}\lab{definitionb}
\eea

where the constant $a$ is defined in \eq{definitionap}. Now
 \eq{EQVN} is the equation for  the BFKL Pomeron fan diagram.
 Fan diagrams 
 can be summed using the generating functional technique that has been developed in $\Lb Y, \zeta\Rb$ representation\footnote{$\zeta$ is conjugate variable to $\nu$ (see more details below)} (see ref. \cite{GENFUN} for details).
 For the generating functional, we can write the linear equation in terms of functional derivatives,
  which reflects the fact that the  Pomeron can decay into two Pomerons. In the dipole model,
  this decay can be written as the decay of one dipole to two dipoles.
   \eq{EQVN} simplifies this functional equation to a recursive formula.
    Two simplification rules are essential for our approach:
   (i) the most singular part of the triple Pomeron vertex has a much simpler form in $\nu$ representation,
       than the BFKL kernel in coordinate representation, and (ii) the loop correction to the vertex can be neglected at high energies
        (see ref. \cite{MPSI} for a full explanation).
        \eq{EQVN}  can be viewed as an equation in time (rapidity). 
Indeed, \eq{EQVN} states that the process of $n$-Pomeron production  can be considered to be the production of $n-1$ Pomerons at time $t$,
and  the later decay of one of the produced Pomerons into two, at time $t+\delta t$  ($\delta t \ll 1$), 
as shown in \fig{vertices1}. Using \eq{GA1} we can rewrite \eq{EQVN} in the following form:

\bea   \label{EQVN1}
&&\hat{\Ga}\Lb n, \la\,\vert\,\{\nu_i\}\Rb \he b \,\sum^{n-1}_{j=1}\hat{\Ga}\Lb n-1,\la\,\vert\,\{\nu_i\}, \nu_j \Rb\,
\Lb\,\frac{\nu_{i}+\nu_{k}}{\nu_{i}\,\nu_{k}}\,\Rb\\
\nn\\
&& i\neq j\neq k\neq l\nn
\eea

where on the LHS of \eq{EQVN1} the notation $\{\nu_i\}=\nu_1,\nu_2,\dots\nu_n$, whereas on the RHS the notation $\{\nu_i\}=\nu_1,\nu_2,\dots\nu_n;\,\,\,\nu_i\neq \nu_j$. 
In this approach
\eq{EQVN1} yields the following solution:

\bea \label{SOL1}
&&\hat{\Ga}\Lb n, \la\,\vert\,\{\nu_i\}\Rb\,\,\,=\,\,\,\,\la \,\Lb n-1\Rb!\,\prod^n_{i=1}\,\Phi\Lb \nu_i\Rb
\,\,\,\,\,\mbox{with}\,\,\,\,\,\,\Phi\Lb \nu_i\Rb\,\,=\,\frac{b}{\nu_i}
 \eea

It is easy to check that this solution satisfies both the recursive equation (see \eq{EQVN1}) and the initial condition of \eq{GA1}.

\subsection{Green function of the resulting BFKL Pomeron}

Using the formula of  \eq{SOL1} for the vertices, we can calculate $\Sigma_n\Lb \omega,\nu\Rb$ from \eq{EQSI} as:

\bea
&&\Sigma_n\Lb \omega, \nu\,=\,i(n - 1)/2 + \lambda\Rb\,\,\,\,=\,\,\,\bas^2\,\Lb -1\Rb^n b^{2n}\frac{ (n-1)!}{n}\,\frac{1}{g(\nu)}\,\Lint\, \prod_i^n \frac{d \nu_i}{2 \pi i} \,\delta\Lb \lambda \,-\, \sum^n_i \nu_i\Rb\,
\frac{1}{\om \,-\,\sum^n_i \om_i(\nu_i)}
\hspace{1cm}\lab{EQSI1}\eea

The explanation behind the factor in front in \eq{EQSI1}, is as follows.
The vertices that enter into \eq{EQSI1}, are equal to  $\hat{\Ga}$ after multiplying by a factor of $
\Lb\om\Lb\nu\Rb-\om\Lb\nu_1\Rb-\om\Lb\nu_2\Rb\Rb/g(\nu)$ (see the definition of \eq{diff1}). In the region where $\la=i/2-\nu\to 0$, this factor reduces to the asymptote
$\bas /\Lb \la\,g(\nu)\Rb$ for large $n$.  
The factor of $
(-1)^n (n - 1)!/n$ that appears in \eq{EQSI1} has the following meaning:

\bea \label{MNG}
(-1)^n (n - 1)!/n\,\,&=&\,\, (-1)^{n-1} \,\mbox{\{ $(-1)$ for each Pomeron loop\}}\,\times\,(n-1)!^2 \,\mbox{\{ from \eq{SOL1} \}}\nn\\
 &\times&\,\,\frac{1}{n!} \,\mbox{\{ from the identity of the Pomerons\}}\,\times (-1)\,\mbox{\{ from definition of $\Sigma$\}}\eea

We do not need to integrate over the entire phase space in $\nu_i$, due to the identity  of Pomerons.
 It is enough to integrate within the region $\nu_n \,> \,\nu_{n-1}\,>\,\dots
\,>\,\nu_i\,\dots\,>\nu_1$.  As one can see, in this region all of the Pomerons have different $
\nu$'s, and can be considered to be different particles. This region covers the $1/n!$ part of the entire phase space in $\nu$.
Further summation  turns out to be simpler in  $Y$ and $\ln \zeta = \ln xx^*$ representation, where $Y$ is the rapidity of the dipole-dipole scattering, while 

\bea
\zeta\he x x^*\he \frac{r^2 R^2}{\Lb \vec{b} + \h(\vec{r} - \vec{R})\Rb^2\,\Lb\vec{b} -  \h(\vec{r} - \vec{R})\Rb^2}\lab{definitionzeta}
\eea

where $\vec{b}$ is the impact parameter of the dipole-dipole scattering, and $\vec{r}$ and $\vec{R}$ are the sizes of the two dipoles. Using these variables, $\Sigma_n\Lb Y, \zeta\Rb$ 
can be calculated using the following transform:

\beq\label{SIYX}
\Sigma_n\Lb Y,\zeta\Rb \,\,=\,\,\Lint^{a + i \infty}_{a - i \infty}\frac{d \omega}{2\pi i} \,e^{\omega Y}\,
\Lint^{i a\prm + \infty}_{ia\prm -\infty}\frac{d \nu }{2\pi } \,e^{i \nu \ln\zeta}\,\,\Sigma_n\Lb \om, \nu\Rb
\eeq

First we  switch to $Y$ representation using the following inverse Mellin transform:

\bea 
\Sigma_n\Lb Y, \nu\Rb \,\,&=&\,\,\Lint^{a + i \infty}_{a - i \infty}\frac{d \omega}{2\pi i} e^{\omega Y} \Sigma_n\Lb \om,\nu\Rb\,\,=\,\,\bas^2\,
(-1)^n\,b^{2n}\frac{ (n-1)!}{n}\,\frac{1}{g(\nu)}\,\Lint \prod_i^n \frac{ d \nu_i}{2 \pi i}\,e^{\omega(\nu_i)\,Y} \,\delta\Lb \lambda \,-\, \sum^n_i \nu_i\Rb
\label{SIYNU}
\hspace{1cm}\eea

Then we switch to $\zeta$ representation using the following approach:

\bea&&
\Sigma_n\Lb Y, \zeta\Rb\he \Lint^{i a\prm + \infty}_{ia\prm -  \infty}\frac{d \nu }{2\pi } \,e^{i \nu \ln\zeta}\,\,\Sigma_n\Lb Y, \nu\Rb\nn\\
\nn\\
&&\He \bas^2\, (-1)^nb^{2n}\,\frac{ (n-1)!}{n}\,\frac{1}{g(\nu)} \Lint^{i a\prm + \infty}_{ia\prm -  \infty}\frac{d \nu }{2\pi } \,e^{i \nu \ln\zeta}
\Lint \prod_i^n \frac{ d \nu_i}{2 \pi i}\,e^{\omega(\nu_i)\,Y} \,\delta\Lb \lambda \,-\, \sum^n_i \nu_i\Rb\lab{lwit}
\eea

As we have discussed generally speaking $\nu \,=\,\nu_n \,-\, \lambda$\footnote{We recall that $\lambda = \ln \zeta/\bas Y \,\ll\,1$}. However, the value of 
$\nu_n$ turns out to be different from $(n-1)!$ in general, for various different $1\to n$ vertex diagrams. In Table 1
 we give the examples for $1\to n$ vertex diagrams, up to $n=6$.

\begin{table}
\begin{center}
\begin{tabular}{|| c | c| c| c| c| c||}
\hline
$\nu_n{\Big/}\Ga$ & $1 \to 2$ & $1 \to 3$ &   $1 \to 4$&  $1 \to 5$&  $1 \to 6$\\
\hline
i & 0 & 0& 0& 4 & 0 \\
\hline
i/2 & 1& 0 & 4&  0 & 80\\
\hline
0 & 0 & 2&0 & 16& 0 \\
\hline
-i/2& 0 &0&2 & 4& 40 \\
\hline
\end{tabular}
\end{center}
\caption{The number diagram for the values of $\nu_n$ for diffrent vetices $\Gamma$.}
\end{table}
 
Unfortunately, we have not
 derived the general rules for how to calculate the value of $\nu_n$ for the $1\to n$ vertex.
We consider two models for such numbers: (i)  each $1\to n$ vertex has $\nu_n=0$, and (ii)
 each $1\to n$ vertex has $\nu_n =  i (n - 1)/2$. 
The first model gives the sum of the leading twist contribution $\zeta^{i \nu} \to 1$,
 while the second model sums over all high twists $\zeta^{i \nu} \to \zeta^{-n/2}$.
  We believe that considering these two models for finding the value of $\nu_n$, provides the largest possible contributions. 
This belief is based on the  following simple examples. As can be seen from \eq{lwit}, we are summing an asymptotic series of the form:

 \beq \label{sum}
 \sum^\infty_{n=0} (- 1)^n\, C_n\, L^n
 \eeq

 where $L$ is a large parameter. Our first  model means that for the leading twist contribution, we choose  $\nu_n=0$ in all $n!$-diagrams.
This leads to $C_n \propto n!$. In the exact approach, the number of  diagrams with $\nu_n=0$ is less than $n!$. 
However, the largest sum corresponds to $C_n = n!$. One can see this by setting $C_n=1$ and $C_n = 1/n!$ in \eq{sum}.
 The same occurs in the second model, which we believe leads to the maximal sum of the highest twist contributions.

\subsection{Summing high twists}
Recall that $\nu\he  i (n-1)/2+\la  \approx i(n-1)/2$  at large $n$\footnote{We recall that $\lambda = \ln \zeta/\bas Y \,\ll\,1$}, such that after switching to the integration variable $\la$, then \eq{lwit} can be written as:

\bea
&&
\Sigma_n\Lb Y, \zeta\Rb\,\,=\,\,\zeta^{(1-n)/2}\,\frac{\bas^2}{4}\,(n-1)^2\,b^{2n}\,(-1)^{n-1}\,\frac{(n-1)!}{n}\,\Lint^{ia\prm+\infty}_{ia\prm-\infty} \f{d \lambda}{2\pi }
\,e^{i \lambda \ln \zeta}\,\Lint \prod_{i=1}^n \frac{ d \nu_i}{2 \pi i}\,e^{\omega(\nu_i)\,Y} \,\delta\Lb \lambda \,-\, \sum^n_i \nu_i\Rb\nn\\
&&= \, \,\zeta^{(1-n)/2}\,\frac{\bas^2}{4}\,(n-1)^2\,b^{2n}\,(-1)^{n-1}\,\frac{(n-1)!}{n}\,\Lint
\prod^n_{i=1}\frac{ d \nu_i}{2 \pi i}\,\,
\, e^{\om(\nu_i) \,Y + i\,\nu_i \ln \zeta}  \lab{SIYNUa}
\eea

where we took the integral over $\lambda$ using the $\delta$-function.  Using \eq{OMNU0}
we can solve the integral over $\nu_i$ explicitly, using the method of steepest descents.
In  this approach \eq{SIYNUa} simplifies to:

\bea \label{SIYNU1}
&&\Sigma_n\Lb Y,\zeta\Rb \he\zeta^{(1-n)/2}\,\frac{\bas^2}{4}\,\, b^{2n}\,(-1)^{n-1}\,\frac{(n-1)^2}{n}(n-1)!\,
\, \Lb \sqrt{\frac{\pi}{ D Y}}\,e^{\om(0)Y - \ln^2\zeta/(4 D Y)}\Rb^n \\
&&\He
\frac{\bas^2}{4}\,\sqrt{\zeta}\, \,\, (-1)^{n-1}\,\frac{(n-1)^2}{n}\,\Gamma\Lb n \Rb \,L^{n}\,\lab{SIYNU2}\\
\nn\\
\mbox{where}\hspace{0.5cm}
&&L\he b^2\,\, \sqrt{\frac{\pi\, }{D\,Y\zeta }} \,\exp\Lb \omega(0)\,Y\,\,-\,\,\frac{\ln^2 \zeta}{4\,D\,Y} \Rb
\lab{L}\eea

Using the integral representation for the Euler-Gamma function (see formula {\bf 8.310(1)} of ref. \cite{GR}):

\beq \label{GAM}
\Gamma\Lb n + 1\Rb \,=\,\int^\infty_0 \,t^{n}\,e^{-\,t}\,d t
;
\eeq

and since we are summing from $n=2$, (since the first $1\to n$ vertex in the sum is the 
$1\to 2$ vertex), we obtain the following result for $\Sigma\Lb Y,\zeta\Rb$:

\bea \label{SIGMASUM}&&
\Sigma\Lb Y,\zeta \Rb \he\,\sum_{n =2}^\infty\, \Sigma_n\Lb Y, \zeta \Rb\he-\frac{\bas^2}{4}\,\sqrt{\zeta}\,\Lint^\infty_0 e^{-t}\,\frac{d t}{t}\,\sum^\infty_{n=2}\,\,\frac{(n-1)^2}{n}\Lb - t L\Rb^n \nn \\
&&\He\,\,-\,\frac{\bas^2}{4}\,\,\sqrt{\zeta}\,\Lint^\infty_0 \,\frac{d t}{t} \,e^{- t}\,\left\{ \frac{Lt + 2(Lt)^2 - \ln( 1 + Lt)
- 2 Lt \ln(1 + Lt) - (Lt)^2 \ln(1 + Lt)}{(1 + Lt)^2}\,\right\}\nn
\\
  &&\He\,\,-\,\frac{\bas^2}{4}\,\,\sqrt{\zeta}\,  \,\Lint^\infty_0 \,\frac{d T}{T} \,e^{- T/L}\, \frac{T + 2T^2 - \ln( 1 + T)
- 2 T \ln(1 + T) - T^2 \ln(1 + T)}{(1 + T)^2}\nn\\
&&\He\,\,\,\frac{\bas^2}{4}\,\,\sqrt{\zeta}\,\Sigma\Lb L \Rb
 \eea

where $T = L t$. 
For large $L$ we find that

\bea \label{LALBE}
\Sigma\Lb L\Rb\,\,
&\xrightarrow{L \gg 1} &\,\Lint^\infty_0 \,\frac{d T}{T} e^{-T/L}\Lb 2 - \ln T+\mathcal{O}\Lb\f{\ln T}{T}\Rb\Rb\,\,\,\approx\,\,\,\Lint^L_0 \,\frac{d T}{T} \Lb - \ln T\Rb\,\,=\,\,-\,(\ln L)^2/2
\eea

Inserting \eq{LALBE} into \eq{SIGMASUM},
using the  definition for $L$  given in \eq{L},  we see that at large $Y$ and in terms of  $\ln \zeta$, then $\Sigma\Lb Y,\zeta\Rb$ tends to the following simple formula:

\beq \label{SIGMAAS}
 \Sigma\Lb Y,\zeta\Rb \,\,\,=\,\,\,\frac{\bas^2}{8}\,\,\sqrt{\zeta}\, \Lb\, \omega_0 Y \,-\,\frac{\ln^2\zeta}{4 D Y} \,-\,\h \ln Y\,-
\h\ln\zeta\,+\,\ln \Lb b^2 \sqrt{\f{\pi}{ D}}\Rb\Rb^2
 \eeq

Using the following formula to transform to $\Lb\,\om,\nu\,\Rb$ representation:

\bea
\Sigma\Lb\om,\nu\Rb\he\Lint^\infty_0 d\zeta\,\zeta^{\,-1-i\nu }\Lint^\infty_0dY \,e^{-\om Y}\Sigma\Lb Y,\zeta\Rb\he
\Lint^\infty_{-\infty} d\ln\zeta\,e^{\, -i\nu\,\ln \zeta}\Lint^\infty_0dY \,e^{-\om Y}\Sigma\Lb Y,\zeta\Rb\lab{followingtransform}\eea

Note that \eq{followingtransform} is the double Mellin transform from $\Lb Y,\zeta\Rb$ to $\Lb\om,\nu\Rb$ representation.
 This corresponds to the inverse-Mellin transform of \eq{SIYX} which 
was used to transform from $\Lb \om,\nu\Rb$ to $\Lb Y,\zeta\Rb$ representation.
Then inserting \eq{SIGMAAS} into \eq{followingtransform}
for large $Y$, leads to the following equation for $\Sigma\Lb \om, \nu\Rb$:

\beq \label{SIGASON}
\Sigma\Lb \om, \nu\Rb\,\,\xrightarrow{\om \to 0}\,\,\frac{\bas^2}{8}\,\,\frac{1}{\Lb i\nu\,+\, \lh \Rb}\,\f{\Lb 2\,\omega^2_0
\,\,+\,\,\om_0\, \om\,\ln \om\,\,+\,\,\,\mathcal{O}\Lb\om^2\Rb
\,\Rb}{\om^3}
\eeq

where the last term $\mathcal{O}\Lb\om^2\Rb$, labels small terms proportional to $\om^2$. In reality, due to the $\om^3$ term in the denominator of \eq{SIGASON},
then overall this $\mathcal{O}\Lb\om^2\Rb$ term, leads to terms which are singular in powers of just $1/\om$, and hence smaller than the singular terms by powers of $1/\om^2$ and $1/\om^3$
that come before. The Green function of the dressed Pomeron reads:

\beq \label{GDP1}
G\Lb \om, \nu\Rb\,\,=\,\,\frac{1}{\om - \om\Lb \nu \Rb + \Sigma\Lb \om,\nu\Rb}\,\,=\,\,\frac{1}{\om - \om\Lb \nu\Rb \,+\,\ml{\frac{\bas^2}{8}\,\,\frac{1}{\Lb i\nu +  \lh \Rb}\,\Lb\frac{2\,\omega^2_0}{\omega^3}\,\,+\,\,\,\om_0\, \frac{\ln \om}{\om^2}\Rb}
}
\eeq

For small $\om$ and $\nu$, we can neglect the contribution from $G_0^{-1}\Lb \om,\nu\Rb = \om - \om\Lb \nu\Rb$. Therefore, in this limit the Green function of the 
dressed Pomeron is given by the following formula:

\bea \label{0GDP2}
G\Lb \om, \nu\Rb&&\,\, \xrightarrow{\om \to 0}\,\,\f{\ml{\frac{8}{\bas^2}\,\frac{\om^3}{\om_0}}\,\Lb \lh +i\nu  \Rb}{\Lb\,2\, \omega_0 
+\,\,\om\,\ln \om \,\Rb}\\
\nn\\
&&\,\,\xrightarrow{\om \to 0}\,\,\frac{4}{\bas^2}\f{\om^3}{\om^2_0}\,\,\Lb i\nu  + \frac{1}{2}\Rb\lab{GDP2}\eea

where in \eq{GDP2} the second term in the denominator of \eq{0GDP2} is neglected in the limit that $\om\to 0$.
One can see that \eq{GDP2} leads to the function that has no singularity at $\om \to 0$ and therefore, the 
corresponding imaginary part of the scattering amplitude vanishes.
To gain a better understanding, keeping the second term in the denominator of \eq{0GDP2} yields:
 
\beq
\label{GDP3}
G\Lb \om, \nu\Rb\,\,=\,\frac{4}{\bas^2}\f{\om^3}{\om_0^2}\,\,\Lb\h + i\nu \Rb\,\Lb 1 \,\,-\,\,\frac{\om\,\ln\om}{2\om_0}\Rb
\,\,=\,\,G^{(1)}\Lb \om, \nu\Rb\,\,+\,\,G^{(2)}\Lb \om, \nu\Rb
\eeq

Passing to ($Y,\zeta$)-representation we see that the asymptotic behaviour of the Pomeron Green function 
is

\bea \label{ASBE0}
G\Lb Y, \zeta\Rb\hspace{0.3cm}&&\He\Lint ^{ia\prm +\infty}_{i a\prm-\infty}\f{d\nu}{2\pi}\,e^{\,i\nu\ln \zeta}\Lint^{a+i\infty}_{a-i\infty}\f{d\om}{2\pi i}\,e^{\,\om Y}
\,G\Lb \om, \nu\Rb \,\nn\\
\nn\\
&&\He\,\Lint ^{ia\prm +\infty}_{i a\prm-\infty}\f{d\nu}{2\pi}\,e^{\,i\nu\ln \zeta}\Lint^{a+i\infty}_{a-i\infty}\f{d\om}{2\pi i}\,e^{\,\om Y}
\,G^{(2)}\Lb \om, \nu\Rb \nn\\
\nn\\
&&\He-\,\frac{2}{\bas^2\,\om^3_0}\,\,\Lb \lh + \f{\D}{\D\ln\zeta}  \Rb\,\delta\Lb \ln \zeta\Rb
\Lint^{\infty}_{0}\,d \om\, e^{\,\,\om Y}\,\,\om^4 \ln \omega\eea

Closing the contour of integration on the real negative $\omega$-axis, we have:

\bea \label{ASBE}
G\Lb Y, \zeta\Rb\hspace{0.3cm}&&\He
\,
\,\frac{2}{\bas^2\,\om^3_0}\,\,\Lb \lh + \f{\D}{\D\ln\zeta}  \Rb\,\delta\Lb \ln \zeta\Rb
\Lint^{\infty}_{0}\,d \om\, e^{\,-\,\om Y}\,\,\om^4 \,\nn\\
&&\He \,\,\,\,\,\,\frac{2}{\bas^2\,\om^2_0}\,\,\Lb \lh + \f{\D}{\D\ln\zeta}  \Rb\,\delta\Lb \ln \zeta\Rb\,\frac{4!}{Y^5}
\eea

From \eq{ASBE} its clear 
that the Green function leads to the cross section that decreases as $1/Y^5$, and  the character of this asymptotic 
behaviour depends on $\zeta$. The factor that depends on  $\zeta$, in front of \eq{ASBE},
indicates that the main contribution turns out to be a leading  twist contribution.\\

This decrease at ultra high energies is
the most salient result of this paper.
It is well known that the Pomeron calculus in zero transverse dimensions, leads to the Green function of the  Pomeron that decreases with energy, (see refs. \cite{AAJ,ALMC,CLR,CI,BOMU}).
However, in our case the decrease of the cross section is only logarithmic, whereas in the Pomeron calculus in zero transverse dimensions, the cross section falls exponentially at large $Y$.
Therefore, we claim that the BFKL evolution of the sizes of the interacting dipoles, do not lead to a new qualitative effect. 
However, it is interesting that the character of the asymptotic behaviour, crucially depends on the size of the interacting dipole.
  
  \subsection{Summing the leading twist contribution}  
  In the previous subsection we demonstrated that summing the contribution of high twists, leads to 
  the solution which appears to be the leading twist contribution. Therefore, it seems reasonably possible to derive the leading twist contribution. 
  We will do this assuming that $\nu_n = 0$ for every odd $n$ (see Table I).  Recalling that $ g(\nu)\to 16\nu^2$ as $\nu \to 0$, and using the formula $\nu=\nu_n-\la$ then
 we  can re-write \eq{lwit} in the form
  
\bea&&
\Sigma_n\Lb Y, \zeta\Rb\he \Lint^{i a\prm + \infty}_{ia\prm -  \infty}\frac{d \lambda }{2\pi } \,e^{i \lambda \ln\zeta}\,\,\Sigma_n\Lb Y, \lambda\Rb\nn\\
\nn\\
&&\He \bas^2\, (-1)^nb^{2n}\,\frac{ (n-1)!}{n}\ \Lint^{i a\prm + \infty}_{ia\prm -  \infty}\frac{d \lambda }{2\pi } \,\frac{1}{16 \,\lambda^2 }\,e^{i \lambda \ln\zeta}
\Lint \prod_i^n \frac{ d \nu_i}{2 \pi i}\,e^{\omega(\nu_i)\,Y} \,\delta\Lb \lambda \,-\, \sum^n_i \nu_i\Rb\lab{ltwit}
\eea  
  
  Using the following representation for the $\delta $- function, namely,

 \beq  \label{DELFU}
 \delta\Lb \lambda \,-\,\sum^n_{i=1} \nu_i\Rb\,\,=\,\,\frac{1}{2 \pi} \Lint^{ \infty}_{- \infty}  \,d \mu\,e^{i \mu\,\Lb \lambda \,-\,\sum^n_{i=1} \nu_i\Rb } 
\eeq

and integrating over $\lambda$ by closing the integration contour around the double pole at $\lambda = 0$, we obtain
  
 \bea \label{LT1}&&
  \Sigma_n\Lb Y, \zeta\Rb \he  \f{ \bas^2}{16}\, (-1)^n\,b^{2n}\,\frac{ (n-1)!}{n}\,\,\frac{1}{2 \pi} \Lint^{ \infty}_{- \infty}  \,d \mu\, \Lb i \ln \zeta + i \mu\Rb 
\prod^n_{i=1} \Lint  \frac{ d \nu_i}{2 \pi i}\,e^{- i \nu_i\,\mu\,\,+\, \,\omega(\nu_i)\,Y } \nn\\
\eea

Integrating over $\nu_i$ using the method of steepest descents, we have:

\bea\lab{0.LT2}
   &&\Sigma_n\Lb Y, \zeta\Rb \he\f{ \bas^2}{16}\, (-1)^nb^{2n}\,\frac{ (n-1)!}{n}\,\,\frac{1}{2 \pi} \Lint^{\infty}_{- \infty}  \,d \mu\, \Lb i \ln \zeta + i \mu\Rb \Lb
 \sqrt{\frac{\pi}{ D\,Y}}\,\exp\Lb \om_0 Y \,\,-\,\,\ml{\frac{\mu^2}{4 D Y}}\Rb\Rb^n \eea

Finally after solving the $\mu$ integral, using the result that $\Lint^\infty_{-\infty} dx \exp\Lb -ax^2\Rb\he \sqrt{\pi/a}$ yields:
\bea
&&\Sigma_n\Lb Y, \zeta\Rb  \he \f{\bas^2}{16}\,\ln \zeta\, (-1)^n\,\,\frac{ (n-1)!}{n}\sqrt{\frac{D\,Y}{\pi\, n}}\,L^n\lab{0.LTL}\\
\nn\\
&& 
L\he b^2\,\, \sqrt{\frac{\pi\, }{D\,Y }} \,e^{ \omega(0)\,Y}  \label{LTL}
\eea

 Using \eq{GAM}  we can re-write \eq{LTL} in the form
   
\bea \label{LT3}
      \Sigma\Lb Y, \zeta\Rb & \he& \f{ \bas^2}{16}\,\sqrt{\frac{D\,Y}{\pi}}\,\ln \zeta\,\sum^\infty_{n=2}\,(-1)^n\,\,\frac{ (n-1)!}{n^{3/2}}\,\,L^n   \,\nn\\
      &=&\,\,\f{ \bas^2}{16}\,\sqrt{\frac{D\,Y}{\pi}}\,\ln \zeta\,\Lint^\infty_{0} d t \,e^{-t} \sum^\infty_{n=2}\,(-1)^n\,\,\frac{ L^n t^{n-1}}{n^{3/2}}\,\, \nn\\
      &=&  \f{\bas^2}{16}\,\sqrt{\frac{D\,Y}{\pi}}\,\ln \zeta\,\Lint^\infty_{0} \frac{d t }{t}\,e^{-t} \,\Big( L t   \,\,+\,\,Li_{3/2}\Lb - L t\Rb\Big)  \eea
   
For properties of the Polylogarithm function see Ref.\cite{LI}. 
Since we  assumed that $\nu_n=0$ occurs only at odd $n$, we can extract this  from the sum of \eq{LT3},
by subtracting from it the function where $L \to -L$. In this approach, finally we arrive at the expression:

   \bea&&
     \Sigma\Lb Y, \zeta \Rb \he \f{ \bas^2}{16}\,\sqrt{\frac{D\,Y}{\pi}}\,\ln \zeta\,\Lb\, 2 L\,+\,\Lint^\infty_{0}
\frac{d T}{T}\,\exp\Lb -T/L\Rb\,\Lb \,Li_{3/2}\Lb - T\Rb  \,\,-\,\,\,Li_{3/2}\Lb T\Rb\Rb\,\Rb\,\,\lab{LT4}\\
\nn\\
&&=\,\,\f{\bas^2}{16}\,\sqrt{\frac{D\,Y}{\pi}}\,\ln \zeta\,\Sigma\Lb L \Rb\lab{LT4b}\eea

where $T= L t$. Since the asymptotic  behaviour of the Polylogarithm function $Li_s(-T) $ is known, namely

\beq \label{ASLI}
Li_s\Lb - T\Rb \,\,\xrightarrow{ T \gg 1}\,\,-\frac{\ln^{s}\Lb T\Rb}{\Gamma(1 + s)} \, 
 \eeq

Then with this in mind, the integral of \eq{LT4} is expected to lead to the following result:
 
\bea \label{LT4a}&&
\Sigma\Lb L\Rb\he2 L + \Lint^\infty_0\frac{d T}{T} \,\exp\Lb-T/L\Rb\,\Lb \,Li_{3/2}\Lb - T\Rb  \,\,-\,\,\,Li_{3/2}\Lb T\Rb\Rb\,\nn\\
\nn\\
&&\xrightarrow{L \gg 1}\,2 L - \Lint^\infty_0\frac{d T}{T} \,\exp\Lb-T/L\Rb \,\frac{\Lb \ln T\Rb^{3/2}}{\Gamma(5/2)}\,\nn\\
\nn\\&&
\He 2 L \,\,-\,\,\,\frac{2}{5}\frac{\Lb \ln L\Rb^{5/2}}{\Gamma(5/2)}\,
\eea

Therefore at large $Y$, after inserting \eq{LT4a} into \eq{LT4b}, yields the following high energy behavior for  
$ \Sigma\Lb Y, \zeta \Rb$:

\beq\label{LT5}
   \Sigma\Lb Y, \zeta \Rb \he \f{\bas^2}{16}\,\,\ln \zeta\,\Lb\, 2
b^2
 e^{\om_0 Y}\,\,-\,\,\frac{1}{\Ga\Lb 7/2\Rb}\sqrt{\f{DY}{\pi}}\Lb\,\, \om_0 Y - \h \ln Y +\ln\Lb b^2\sqrt{\f{\pi}{D}}\Rb\,\,\Rb^{5/2}\,\Rb
   \eeq

Using the formula of \eq{followingtransform} to switch to $\Lb\om,\nu\Rb$ representation, by approximating the second term in \eq{LT5}
at $Y$ as $\sqrt{\ml{\f{DY}{\pi}}}\Lb \om_0 Y - \lh \ln Y +\ln\Lb b^2\sqrt{\ml{\f{\pi}{D}}}\Rb\Rb^{5/2}\approx
\ml{\sqrt{\f{D}{\pi}}}\Lb \om_0^{5/2}Y^3-\ml{\f{5}{4}}\ln Y\om_0^{3/2}Y^2+\mathcal{O}\Lb Y\Rb\Rb$, leads to the formula:

 \beq\label{LT6}
\Sigma\Lb \om, \nu\Rb\,\, \he \,\f{\bas^2}{16}\,\,\frac{1}{\nu^2}\,\Lb\,- \f{2b^2}{\om \,-\,\om_0}\,+\,\frac{1}{\Gamma\Lb 7/2\Rb }
\sqrt{\f{D}{\pi}}\Lb 6\f{\om_0^{5/2}}{\om^4} \,\,+\,\,\frac{5}{2}\frac{\om_0^{3/2}}{\om^3}\,\Lb \ln \om\,\,+\,\,{\cal O}(\mbox{Const})\Rb\,\Rb\,\Rb  
\eeq

where $\mathcal{O}\Lb\mbox{Const}\Rb$ denotes constant terms that do not depend on $\om$. 
Plugging  \eq{LT6} into the formula for the  dressed Pomeron propagator $G_2\Lb \om,\nu\Rb$, defined in \eq{1.G2},
then  one arrives at the following expression:

\bea \label{LT7}
&&G_2\Lb \om, \nu\Rb\he\frac{1}{\om - \om\Lb \nu \Rb + \Sigma\Lb \om,\nu\Rb}\\
&&\He\frac{1}{\om - \om\Lb \nu\Rb \,+\,\,\ml{\f{ \bas^2}{16}}\,
\,\ml{\frac{1}{\nu^2}}\,\ml{\Lb\, -\frac{2b^2}{\om \,-\,\om_0}\,+\,\frac{1}{\Gamma\Lb 7/2\Rb }
\ml{\sqrt{\f{D}{\pi}}}\,\Lb 6\f{\om_0^{5/2}}{\om^4} \,\,+\,\,\frac{5}{2}\frac{\om_0^{3/2}}{\om^3}\, \ln \om\,\Rb\,\Rb } }\lab{LT7a}\eea

From inspection of Eqs. (\ref{LT6}) and (\ref{LT7}), then $\om \to \om_0$ leads to the asymptote  $G_2\Lb \om, \nu\Rb\,\propto\,\om - \om_0$
 and therefore, this region does not contribute to the cross section due to the absence of singularities in the region $\om = \om_0$.
 In the limit that $\om \to 0$, \eq{LT7a} tends to the following limit (where the third term in the denominator of \eq{LT7a} dominates in this region):
 
\bea
G_2\Lb \om, \nu\Rb\hspace{0.3cm}
&&\xrightarrow{\om\,\to\,0}\hspace{0.3cm} \frac{1}{\,\,\ml{\f{ \bas^2}{16}}\,
\,\ml{\frac{1}{\nu^2}}\,\ml{\Lb\, \,\frac{1}{\Gamma\Lb 7/2\Rb }
\ml{\sqrt{\f{D}{\pi}}}\,\Lb 6\f{\om_0^{5/2}}{\om^4} \,\,+\,\,\frac{5}{2}\frac{\om_0^{3/2}}{\om^3}\, \ln \om\,\Rb\,\Rb } }
\nn\\\nn\\
&&\He\frac{16 \,\nu^2}{\bas^2}\sqrt{\frac{\pi}{D}}\,\Ga\Lb\f{7}{2}\Rb \,\frac{\om^4}{6 \om_0^{5/2}}\,\Lb 1\,\,-\,\,\f{5}{12}\,\frac{\om}{\om_0}\,\ln \om\,+\,
\mathcal{O}\Lb \om^2\ln^2\om\Rb\Rb                     
\hspace{1cm}\lab{LT7b}\eea

The first term in brackets in \eq{LT7b} leads to  $G_2\Lb \om, \nu\Rb\,\propto\,\om^4$,
which also does not contribute to the total cross section, thanks to the absence of any singularities in this region.
Hence it follows that the first contribution stems from the second term in \eq{LT7b},
which
leads to the contribution to  $G_2\Lb \om ,\nu\Rb$ equal to:

\bea
&&G_2\Lb\om,\nu\Rb
\he -\f{25\pi\nu^2}{12\,D^{\h}\bas^2} \,\, \,\frac{\om^5}{\om_0^{7/2}}\,\,\ln \om\,\lab{LT7c}\eea

where in passing from \eq{LT7b} to \eq{LT7c} the fact that $\Ga\Lb 7/2\Rb\,=\,15\pi^{1/2}/8$ was used. Finally 
passing to ($Y,\zeta$)-representation we see that the asymptotic behaviour of the Pomeron Green function 
is

\bea \label{LT8}&&
G_2\Lb Y, \zeta\Rb\he\Lint ^{ia\prm +\infty}_{i a\prm-\infty}\f{d\nu}{2\pi}\,e^{\,i\nu\ln \zeta}\Lint^{a+i\infty}_{a-i\infty}\f{d\om}{2\pi i}\,e^{\,\om Y}
\,G_2\Lb \om, \nu\Rb \,\nn\\
\nn\\
&&\He\f{25\pi}{12\,D^{\h}\bas^2\,\om_0^{7/2}} \,\,
\f{\D^2}{\D\ln\zeta^2}\,\de\Lb \ln \zeta \Rb\,\frac{1}{Y^6}
\eea

Hence its clear from \eq{LT8} that the summation of the leading twist contribution, leads to the same qualitative result, namely that the  Pomeron Green function 
vanishes at large $Y$, but only logarithmically.
This style of decrease is steep enough to provide the final answer, without the need for further re-summation.

It should be mentioned that  in spite of the decreasing behavior  of the amplitude with  energy, at large values of $b$  the integral over the 
impact parameter turns out to be divergent, indicating that the problem of the large $b$-dependence
cannot not be cured by summing enhanced diagrams.
Moreover, this problem needs new ideas from non-perturbative QCD, in order to find a solution.

 \section{Conclusions}
This paper describes the technique developed to find the sum of enhanced diagrams ( Pomeron loops), in the dipole-dipole scattering process. 
In conclusion we would like to mention
two main features of the result. The first one, that the cross section and/or the Green function of the  dressed BFKL Pomeron falls down with energy. The second result,
 is that 
the  asymptotic  behaviour depends  crucially on the size of the colliding dipoles, and  the impact parameter of the collision.

We wish also to draw the attention of our reader, to two selection rules which are essential for our approach to the summation of 
the enhanced diagrams. First, we restrict ourselves to the contribution to the triple BFKL Pomeron vertex (see \eq{tpv}), that is singular in the region 
$i/2-\nu-\nu_1-\nu_2\to 0$, for small $\nu_1$ and $\nu_2$,
 since this part of the vertex generates the most singular contribution, leading to a larger result than the other parts of the vertex.
  Second, we neglected the contribution from Pomeron loops to the dressed vertices. This assumption
 is equivalent to the Mueller-Patel-Salam-Iancu approximation \cite{MPSI}, formulated in the $s$-channel of the reaction.
 Therefore, we sum the Pomeron loop diagrams, but using the above mentioned specific assumption 
 about Pomeron vertices.  It should be stressed that the Mueller-Patel-Salam-Iancu approximation, as well as our approximation,
 selects the diagrams with the most essential increase with energy, and therefore, it  can be used in our approach.

As we have discussed in the introduction, we cannot prove the BFKL Pomeron calculus, based only on the triple Pomeron vertices. Moreover, we personally
have an argument stating, that it is necessary to introduce the four BFKL Pomeron vertex \cite{KLP}.
The fact that we obtained the total cross section that decreases with energy,
 stands as a reminder 
 of the constant  cross section obtained, only after taking into account the four Pomeron vertex, in $1+1$ dimensional Pomeron calculus.  However, we would like to stress that in our case, there is only a logarithmic decrease, which is steep enough to
 claim that the summation of the enhanced diagram provides the solution to the problem.
 
 As we have mentioned in the introduction, using the BFKL Pomeron calculus in the form of \eq{BFKLFI}, we neglected both the vertices of transition of one
 Pomeron to more than two Pomerons, as well as the contribution of the multi -gluon states in the next-to-leading $1/N_c$ order (see \fig{mpom}).
 
\FIGURE[h]{\begin{minipage}{140mm}{
\centerline{\epsfig{file=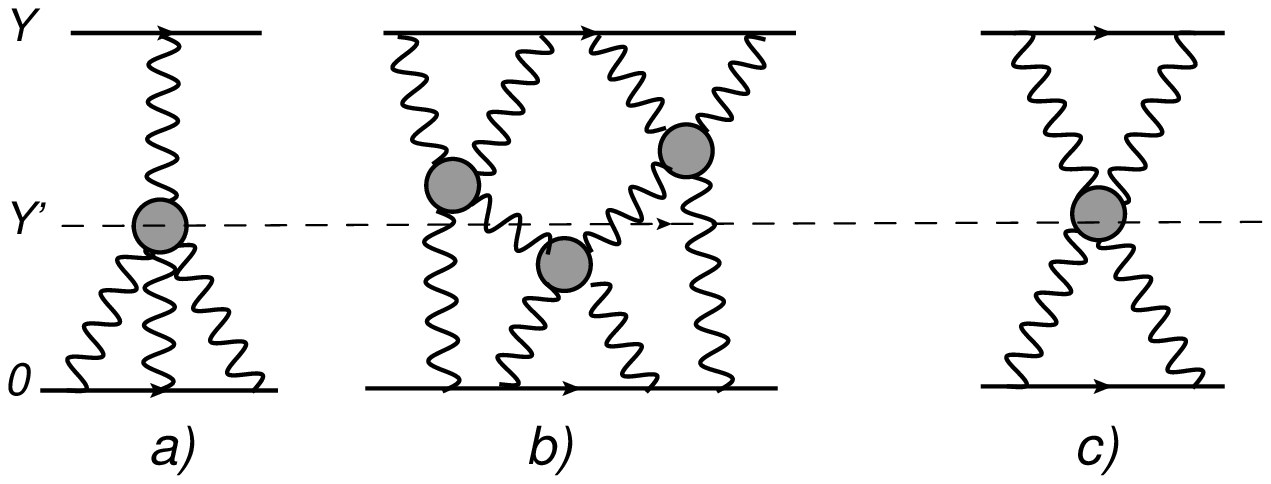,width=120mm}}}
\end{minipage}
\caption{ The vertex of one Pomeron to three Pomerons (\fig{mpom}-a); the contribution of the multi-Pomeron states (\fig{mpom}-b)
 and the first correction to the two Pomerons state (\fig{mpom}-c). The wavy lines denote the Green function of the dressed Pomeron.}
 \label{mpom}}

 In \fig{mpom} the wavy lines denote the dressed Pomeron Green function ($G$),
which is proportional to $Y^{-5}$ or to $Y^{-6}$ depending on the model.
Therefore, the contribution of the diagram of \fig{mpom}-a is equal to

 \bea \label{V13}
 A_{13}\, \,&\propto&\,\int d Y'\,\Gamma\Lb 1\to 3\Rb G\Lb Y - Y'\Rb\,G^3\Lb Y'\Rb\\
 & \rightarrow&\,\Gamma\Lb 1 \to 3\Rb \int d Y' \Lb Y - Y'\Rb^{-5}\,Y'^{-15} \propto \Gamma\Lb 1 \to 3\Rb Y^{-5} \propto G\Lb Y\Rb\nn
  \eea

 One can see that the new vertex $ \Gamma\Lb 1 \to 3\Rb$ led to the renormalization of the coupling of the Pomeron with the dipole, that is included in \eq{BFKLFI} in the term $S_E$.  In other words, the new vertex did not change the BFKL Pomeron calculus.

 The multi-gluon state, i.e.  2$n$-gluons   in the $t$ channel,  generates  a  larger intercept,  than
 the intercept that  the corresponding number ($n$) Pomerons would generate. As  shown in Ref. \cite{MPOM},
 the new intercept can be calculated by summing the Pomeron exchanges, with the new vertex of the interaction of two Pomerons (see \fig{mpom}-b).
  The contribution of the first diagram shown in \fig{mpom}-c is equal to
 \bea \label{V22}
 A_{22}\,& \propto & \,\int d Y'\,\Gamma\Lb 2 \to 2\Rb G^2\Lb Y - Y'\Rb\,G^2\Lb Y'\Rb\\
 & \rightarrow&\,\Gamma\Lb 2 \to 2\Rb \int d Y' \Lb Y - Y' \Rb^{-10}\,Y'^{-10} \propto \Gamma\Lb 2\to 2\Rb Y^{-10} \propto G^2\Lb Y\Rb\,\ll\,G\Lb Y\Rb\nn
 \eea

 Therefore, the contribution of the multi -Pomeron exchanges with the interaction between them leads to small, negligible contributions. 
 
 In light of this, we conclude that the correction to the BFKL Pomeron calculus cannot change the fact that the total cross section falls at high energy.
 Of course, we have discussed only those multi-gluon states in the $t$ - channel, 
where each pair of gluons is in a colourless state. These states have been discussed in Ref.\cite{MPOM}. As far as we know, this is the only known example
of the source of the intercept, which is larger than the intercept of the exchange of several Pomerons.  Unfortunately, we do not know
about  the situation with the exchange of more general $n$-gluon states, which could make the BFKL Pomeron calculus  incorrect in the $1/N_c$ approximation.However, the examples of such states have not been found ( see Ref.
 \cite{KOKOMA}).
 
 We would like to recall, that the final re-summation was applied in the case of two extreme models, for the $\nu$ dependence of the 1 Pomeron $\to$ $n$ Pomerons transition
 vertices. The answer to this problem, is the next challenging problem which we hope to resolve 
 in the near future.

 Nevertheless, we believe that the problem addressed in this paper, is the first  to be solved in order
 to start  a theoretical discussion on what could happen with the  dilute-dilute system of scattering, at high energy.
Observing the substantial  difference between the Pomeron self-energy and its Green function, 
we conclude that the quantum effects due to the BFKL Pomeron interaction, change the character of the asymptotic behaviour.
Since these effects are so strong,
  they have to be taken into account for dilute-dense and dense-dense scattering, at ultra high energies.

\section{Acknowledgements}

This research was supported by the Funda\c{c}$\tilde{a}$o para ci$\acute{e}$ncia e a tecnologia (FCT),
and CENTRA - Instituto Superior T$\acute{e}$cnico (IST), Lisbon and  by the  Fondecyt (Chile) grant 1100648. 
One of us (JM) would like to thank Tel Aviv University for their hospitality on this visit,
during the time of the writing of this paper.

\renewcommand{\theequation}{A.\arabic{equation}}
\setcounter{equation}{0}  
\begin{boldmath}
\section{Appendix A - The BFKL kernel and the triple Pomeron vertex} \lab{seca}
\end{boldmath}

The BFKL kernel $\om\Lb\nu\Rb$ is defined as

\bea
&&\om\Lb\nu\Rb\hspace{0.3cm}=\hspace{0.3cm}\bas\chi\Lb\nu\Rb\,;\hspace{3cm}\,\Lb\,\bas\hspace{0.3cm}=\hspace{0.3cm}\f{\as N_c}{\pi}\,\Rb\lab{bfkl}\\
\nn\\
&&\chi\Lb\nu\Rb\hspace{0.3cm}=\hspace{0.3cm}\Re\Lb\psi\Lb 1\Rb-\psi\Lb \h+i\nu\Rb\Rb\hspace{0.3cm}=\hspace{0.3cm}2\psi\Lb 1\Rb-\psi\Lb\h+i\nu\Rb-\psi\Lb\h-i\nu\Rb\lab{chi}
\eea

where $\psi\Lb x\Rb$ is the Di-Gamma function, defined as  \cite{GR}:

\bea
\psi\Lb x\Rb\hspace{0.3cm}&&=\hspace{0.3cm}\f{d}{dx}\ln \Ga\Lb x\Rb\hspace{0.3cm}=\hspace{0.3cm}-\,\ga_E-\sum^\infty_{k=0}\Lb \f{1}{x+k}-\f{1}{1+k}\Rb\lab{psi}\eea

In the vicinity of $\nu \to 0$, the BFKL kernel takes the following form
\beq \label{OMNU0}
\om\Lb \nu\Rb\hspace{0.3cm}= \,\,\,\,4\bas\,\ln 2-14\bas\zeta\Lb 3\Rb \nu^2\,\,=\,\, \omega\Lb 0\Rb \,\,-\,\,D \nu^2\,\,\equiv\,\,\om_0\,\,-\,\,D \nu^2
\eeq

{\bf \large The planar vertex}

The triple Pomeron vertex $\Gamma\Lb\nu\,\vert\,\nu_1,\nu_2\Rb$ was defined above in \eq{tpv},
in terms of two contributing diagrams, namely the planar and non-planar diagrams shown in \fig{fvertex}.
The expression for the planar diagram given in \eq{planar}, contained the function
$\Om\Lb\nu\vert\nu_1,\nu_2\Rb$, which is given by the following integral (for a detailed explanation and derivation of these results, see
Ref. \cite{1.Korchemsky:1997fy}):

\bea
&&\Om\Lb\nu\vert\nu_1,\nu_2\Rb=\!\!\int\f{d^2z_0d^2z_1d^2z_2}{\mid z_{01}z_{12}z_{20}\mid^2}\!\!
\Lb\f{z_{01}}{z_0z_1} \f{\bar{z_{01}}}{\bar{z}_0\bar{z}_1}\Rb^{\!\lh+i\nu}\!\!
\!\Lb\f{z_{12}}{\Lb 1-z_1\Rb\Lb 1-z_2\Rb}\!\f{\bar{z}_{12}}{\Lb 1-\bar{z}_1\Rb\Lb 1-\bar{z}_2\Rb}\Rb^{\!\lh+i\nu_1}\!\!\!
\biggl( z_{20}\bar{z}_{20}\biggr)^{\!\lh+i\nu_2}\hspace{0.9cm}\lab{2.Omega}\eea

The integrations of \eq{2.Omega} were evaluated in \cite{1.Korchemsky:1997fy}, where the following results were derived. 
Note that in the original paper of ref. \cite{1.Korchemsky:1997fy}, the symmetric property $\Om\Lb\nu_1\,\vert\,\nu_2,\nu_3\Rb\,=\,
\Om\Lb\nu_3\,\vert\,\nu_1,\nu_2\Rb\,=\,\Om\Lb\nu_2\,\vert\,\nu_3,\nu_1\Rb$ was derived. We have used this
symmetric property, and hence the version of the function $\Om$ written below is based on a different permutation of arguments and differs from the version
given in ref. \cite{1.Korchemsky:1997fy}.

\bea
&&\Om\Lb\nu\,\vert\,\nu_1,\nu_2\Rb\hspace{0.3cm}=\hspace{0.3cm}\Om_1\Lb\nu\,\vert\,\nu_1,\nu_2\Rb
 \hspace{0.3cm}+\hspace{0.3cm}\Om_2\Lb\nu\,\vert\,\nu_1,\nu_2\Rb\hspace{0.3cm} +\hspace{0.3cm}
\Om_3\Lb\nu\,\vert\,\nu_1,\nu_2\Rb\lab{thesumofomegas}\\
\nn\\
&&\Om_1\Lb\nu\,\vert\, \nu_1,\nu_2\Rb\hspace{0.3cm}=\hspace{0.3cm}\pi^3\,\,\f{\Ga\biggl( \lh+i\nu_1+i\nu_2-i\nu\biggr)}{\Ga\biggl(\lh-i\nu_1-i\nu_2+i\nu\biggr)}\,\,
\f{\Ga^2\biggl(\lh-i\nu_1\biggr)\Ga^2\biggl(\lh-i\nu_2\biggr)}{\Ga^2\biggl(\lh-i\nu\biggr)}\lab{om1nu}\\
&&\times\,\Lint^1_0 dx\Lb 1-x\Rb^{-\mathlarger{\h}-i\nu}\,\,_2F_1\Lb \h+i\nu_1,\h-i\nu_1\begin{vmatrix} 1\end{vmatrix} x\Rb\,_2F_1\Lb \h+i\nu_2,\h-i\nu_2 \begin{vmatrix} 1\end{vmatrix}x\Rb\nn\\
&&\times\,\Lint^1_0 dy \,y^{-\lh-i\nu}\Lb 1-y\Rb^{-\lh-i\nu_1-i\nu_2+i\nu}\,\,_2F_1\biggl( \h-i\nu_1,\h-i\nu_1   \begin{vmatrix} 1\end{vmatrix} y\biggr)\,_2F_1\biggl( \h-i\nu_2,\h-i\nu_2
\begin{vmatrix} 1\end{vmatrix} y\biggr)\nn\\
\nn\\\nn\\
&&\Om_2\Lb\nu\,\vert\, \nu_1,\nu_2\Rb\hspace{0.3cm}=\hspace{0.3cm}\pi^3\,\,\f{\Ga\biggl( \lh+i\nu_1+i\nu_2-i\nu\biggr)}{\Ga\biggl(\lh-i\nu_1-i\nu_2+i\nu\biggr)}\,\,
\f{\Ga\biggl(\lh+i\nu\biggr)\Ga\biggl(\lh-i\nu\biggr)\Ga\biggl( \lh-i\nu_1\biggr)\Ga^2\biggl( \lh-i\nu_2\biggr)}
{\Ga\biggl(\lh+i\nu_1\biggr)}\hspace{1cm}\lab{om2nu}\\
\nn\\&&\times\,\underline{\,_4F_3\biggl( \lh+i\nu_2,\lh-i\nu_2,\lh-i\nu,\lh-i\nu\begin{vmatrix} 1,1+i\nu_1-i\nu,1-i\nu_1-i\nu\end{vmatrix} 1\biggr)}\nn\\
&&\hspace{3cm}\Ga\biggl(1+i\nu_1-i\nu\biggr)\Ga\biggl(1-i\nu_1-i\nu\biggr)\nn\\
\nn\\
&&\times\,\Lint^1_0 dx\,\Lb 1-x\Rb^{-\lh-i\nu_1-i\nu_2+i\nu}\,\,_2F_1\biggl(  \lh-i\nu_1,\lh-i\nu_1\begin{vmatrix} 1\end{vmatrix}x\biggr)\,_2F_1\biggl( \lh-i\nu_2,\lh-i\nu_2  \begin{vmatrix} 1\end{vmatrix} x \biggr)\nn\\
\nn\\\nn\\
&&\Om_3\Lb\nu\,\vert\, \nu_1,\nu_2\Rb\hspace{0.3cm}=\hspace{0.3cm}\Om_2\Lb \nu\,\vert\,\nu_2,\nu_1\Rb\hspace{0.5cm}\lab{om3nu}\eea

Using the following identity for the Hyper-geometric function (see Ref.\cite{GR} formula {\bf 9.100})

\bea
\,_2F_1\Lb\,a,b\,\begin{vmatrix}c\end{vmatrix}x\Rb\he\Lb 1-x\Rb^{c-a-b}\,_2F_1\Lb c-a,c-b\begin{vmatrix}c\end{vmatrix}x\Rb\lab{1.2F1}
\eea

then the factor $\,_2F_1\Lb\lh-i\nu_1,\lh-i\nu_1\begin{vmatrix}1\end{vmatrix}x\Rb 
\,_2F_1\Lb\lh-i\nu_2,\lh-i\nu_2\begin{vmatrix}1\end{vmatrix}x\Rb$ that appears in Eqs. (\ref{om1nu}) and (\ref{om2nu}) may be re-written, such that:

\bea
&&\Om_1\Lb\nu\,\vert\, \nu_1,\nu_2\Rb\he
\pi^3\,\,\f{\Ga\biggl( \lh+i\nu_1+i\nu_2-i\nu\biggr)}{\Ga\biggl(\lh-i\nu_1-i\nu_2+i\nu\biggr)}\,\,
\f{\Ga^2\biggl(\lh-i\nu_1\biggr)\Ga^2\biggl(\lh-i\nu_2\biggr)}{\Ga^2\biggl(\lh-i\nu\biggr)}\lab{1.om1nu}\\
&&\times\,\Lint^1_0 dx\Lb 1-x\Rb^{-\mathlarger{\h}-i\nu}\,\,_2F_1\Lb \h+i\nu_1,\h-i\nu_1\begin{vmatrix} 1\end{vmatrix} x\Rb\,_2F_1\Lb \h+i\nu_2,\h-i\nu_2 \begin{vmatrix} 1\end{vmatrix}x\Rb\nn\\
&&\times\,\Lint^1_0 dy \,y^{-\lh-i\nu}\Lb 1-y\Rb^{-\lh + i\nu_1+ i\nu_2+i\nu}\,\,_2F_1\biggl( \h + i\nu_1,\h + i\nu_1   \begin{vmatrix} 1\end{vmatrix} y\biggr)
\,_2F_1\biggl( \h + i\nu_2,\h + i\nu_2
\begin{vmatrix} 1\end{vmatrix} y\biggr)\nn\\
\nn\\\nn\\
&&\Om_2\Lb\nu\,\vert\, \nu_1,\nu_2\Rb\hspace{0.3cm}=\hspace{0.3cm}\pi^3\,\,\f{\Ga\biggl( \lh+i\nu_1+i\nu_2-i\nu\biggr)}{\Ga\biggl(\lh-i\nu_1-i\nu_2+i\nu\biggr)}\,\,
\f{\Ga\biggl(\lh+i\nu\biggr)\Ga\biggl(\lh-i\nu\biggr)\Ga\biggl( \lh-i\nu_1\biggr)\Ga^2\biggl( \lh-i\nu_2\biggr)}
{\Ga\biggl(\lh+i\nu_1\biggr)}\hspace{1cm}\lab{1.om2nu}\\
\nn\\&&\times\,\underline{\,_4F_3\biggl( \lh+i\nu_2,\lh-i\nu_2,\lh-i\nu,\lh-i\nu\begin{vmatrix} 1,1+i\nu_1-i\nu,1-i\nu_1-i\nu\end{vmatrix} 1\biggr)}\nn\\
&&\hspace{3cm}\Ga\biggl(1+i\nu_1-i\nu\biggr)\Ga\biggl(1-i\nu_1-i\nu\biggr)\nn\\
\nn\\
&&\times\,\Lint^1_0 dx\,\Lb 1-x\Rb^{-\lh + i\nu_1+ i\nu_2+i\nu}\,\,_2F_1\biggl(  \lh + i\nu_1,\lh + i\nu_1\begin{vmatrix} 1\end{vmatrix}x\biggr)
\,_2F_1\biggl( \lh +i\nu_2,\lh +i\nu_2  \begin{vmatrix} 1\end{vmatrix} x \biggr)\nn\\
\nn\\\nn\\
&&\Om_3\Lb\nu\,\vert\, \nu_1,\nu_2\Rb\hspace{0.3cm}=\hspace{0.3cm}\Om_2\Lb \nu\,\vert\,\nu_2,\nu_1\Rb\hspace{0.5cm}\lab{1.om3nu}\eea

Using the following expansions for the Hyper-geometric functions  (see Ref.\cite{GR} formulae {\bf 9.100}  and {\bf 9.14}):

\bea&&
\,_2F_1\Lb a,b\,\begin{vmatrix} c\end{vmatrix} x\Rb\he
 \f{\Ga\Lb c\Rb}{\Ga\Lb a\Rb\Ga\Lb b\Rb}\sum^\infty_{n=0} \f{\Ga\Lb a+n\Rb\Ga\Lb b+n\Rb}{\Ga\Lb c+n\Rb}\f{x^n}{n!}\lab{2.2F1}\\
\nn\\
&&
\,_4F_3\Lb a,b,c,d\,\begin{vmatrix} f,g,h\end{vmatrix} x\Rb\he
 \f{\Ga\Lb f\Rb \Ga\Lb g\Rb \Ga\Lb h\Rb}{\Ga\Lb a\Rb\Ga\Lb b\Rb \Ga\Lb c\Rb\Ga\Lb d\Rb}
\sum^\infty_{n=0} \f{\Ga\Lb a+n\Rb\Ga\Lb b+n\Rb \Ga\Lb c+n\Rb\Ga\Lb d+n\Rb}{\Ga\Lb f+n\Rb \Ga\Lb g+n\Rb \Ga\Lb h+n\Rb}\f{x^n}{n!}
\hspace{1.5cm}
\lab{4F3}
\eea

Then Eqs. (\ref{1.om1nu} - \ref{1.om3nu}) can be re-cast in the following form:

\bea
&&\Om_1\Lb\nu\,\vert\, \nu_1,\nu_2\Rb\he 
\pi^3\,\,\f{\Ga\biggl( \lh +i\nu_1+i\nu_2-i\nu\biggr)}{\Ga\biggl(\lh-i\nu_1-i\nu_2+i\nu\biggr)}\,\,\f{\Ga\Lb\lh -i\nu_1\Rb\Ga\Lb\lh -i\nu_2\Rb}{\Ga^2\Lb\lh-i\nu\Rb
\Ga^3\Lb\lh + i\nu_1\Rb
\Ga^3\Lb \lh +i\nu_2\Rb}\lab{2.om1nu}\\
\nn\\\nn\\
&&\ml{\ml{\sum}}^\infty_{m,n=0}\,\,\Ga\Lb\lh +i\nu_1+m\Rb\Ga\Lb\lh -i\nu_1 + m\Rb \Ga\Lb\lh +i\nu_2+n\Rb\Ga\Lb\lh -i\nu_2 + n\Rb\,\,\,
\f{B\Lb m+n+1 \,\vert\,\lh -i\nu\Rb}{\Ga^2\bl 1+m\br\Ga^2\bl 1+n\br }\nn\\
\nn\\
&&\ml{\ml{\sum}}^\infty_{p,r=0}\,\,\Ga^2\Lb\lh  + i\nu_1+p\Rb\Ga^2\Lb\lh  + i\nu_2 + r\Rb\,\,
\f{B\Lb  \lh - i\nu + p + r\,\vert\,\lh + i\nu + i\nu_1 +i\nu_2 \Rb}{\Ga^2\bl 1+p\br\Ga^2\bl 1+r\br }\,\,\nn\eea

\bea
&&\Om_2\Lb\nu\,\vert\, \nu_1,\nu_2\Rb\he
\pi^3\,\,\f{\Ga\biggl( \lh +i\nu_1+i\nu_2-i\nu\biggr)}{\Ga\biggl(\lh-i\nu_1-i\nu_2+i\nu\biggr)}\,\,\f{\Ga\Lb\lh +i\nu\Rb\Ga\Lb\lh -i\nu_1\Rb 
\Ga\Lb\lh -i\nu_2\Rb}{\Ga\Lb\lh-i\nu\Rb
\Ga^3\Lb\lh + i\nu_1\Rb
\Ga^3\Lb \lh +i\nu_2\Rb}\lab{2.om2nu}\\
\nn\\
\nn\\&&\times\hspace{0.3cm}\ml{\ml{\sum^\infty_{p=0}}}\,\,\,
\f{\Ga\Lb\lh +i\nu_2 +p\Rb\Ga\Lb\lh - i\nu_2 +p\Rb\Ga^2\Lb\lh-i\nu +p\Rb}{\Ga\bl 1-i\nu +i\nu_1+p\br\Ga\bl 1-i\nu -i\nu_1+p\br}\f{1}{\Ga^2\bl 1+p\br}\nn\\
\nn\\
&&\times\,\hspace{0.3cm}
\ml{\ml{\sum^\infty_{m,n=0}}}\,\,\,\Ga^2\Lb\lh + i\nu_1 +m\Rb \Ga^2\Lb\lh  + i\nu_2 + n\Rb \f{
B\Lb 1+m+n\,\vert\,
\lh +i\nu + i\nu_1 + i\nu_2\Rb}{\Ga^2\bl
 1+m\br\Ga^2\bl 1+n\br}
\nn\\\nn\\\nn\\
&&\Om_3\Lb\nu\,\vert\, \nu_1,\nu_2\Rb\hspace{0.3cm}=\hspace{0.3cm}\Om_2\Lb \nu\,\vert\,\nu_2,\nu_1\Rb\hspace{0.5cm}\lab{2.om3nu}\eea

where $B\Lb m\,\vert\,n\Rb$ is the well known beta-function, defined as (see Ref.\cite{GR} formula {\bf 8.380}):

\bea&&
B\Lb m\,\vert\,n\Rb\he \Lint^1_0 dx \,x^{m-1}\Lb 1-x\Rb^{n-1}\he\f{\Ga\Lb m\Rb\Ga\Lb n\Rb}{\Ga\Lb m+n\Rb}\lab{beta}\eea

 In the region $\nu_1 \to 0$ and $\nu_2 \to  0$ we can simplify the expressions for the $\Om_i$. 
All of the $\Om_i$ contain a pole at $ i/2-  \nu -\nu_1 - i \nu_2 =0$. In vicinity of this pole: 

\beq \label{B1}
B\Lb n \vert \h + i \nu + i\nu_1 + i \nu_2\Rb \,\,\rightarrow\,\,\frac{1}{ \h + i \nu + i\nu_1 + i \nu_2}\,\,+\mbox{\{non singlular terms\}}
\eeq

Using the limit of \eq{B1} we can calculate the residue of the pole at
 $ i/2-  \nu -\nu_1 - i \nu_2 =0$
of the function $\Om_2\Lb\nu\,\vert\,\nu_1,\nu_2\,\Rb\he \Om_3\Lb\nu\,\vert\,\nu_2,\nu_1\,\Rb$, for the 
even narrower region where  $\nu_1 \to 0$ and $\nu_2 \to  0$. In this approach, from \eq{2.om2nu} in this region,  the residue is:

\bea \label{om2snu}&& \ml{\ml{\lim_{\substack{   i/2-  \nu -\nu_1 - i \nu_2 \to 0 \\ \nu_1,\nu_2\to 0   }}}}\hspace{0.3cm}
\Lb  i/2 - \nu  -\nu_1 -\nu_2\,\Rb\,\Om_2\Lb\nu\,\vert\, \nu_1,\nu_2\Rb\he 4 \pi^3\,\,\,_4F_3\Lb \lh,\lh,1
,1 \begin{vmatrix} 1, \lhh, \lhh\end{vmatrix} 1\Rb\nn\\
\nn\\
&&\times\,\,\,\ml{\ml{\lim_{\substack{ \nu_1,\nu_2\to 0   }}}}\hspace{0.3cm}\,_2F_1\Lb
  \lh + i\nu_1,\lh + i\nu_1\begin{vmatrix} 1\end{vmatrix} 1 \Rb
\,_2F_1\Lb\lh +i\nu_2,\lh +i\nu_2  \begin{vmatrix} 1\end{vmatrix} 1 \Rb\nn\\
\nn\\
\nn\\
&&\He  \ml{\ml{\lim_{\substack{\nu_1,\nu_2\to 0   }}}}\hspace{0.3cm}
-\,\,\f{\pi}{3\nu_1\nu_2} \label{om2snuf}
\eea

The identical result is true for $\Om_3\Lb\nu\,\vert\,\nu_1,\nu_2\Rb$. Repeating this procedure we obtain that $\Om_1\propto \Lb \nu_1 + \nu_2\Rb \ll \Om_2$.
Thus in the region  $ i/2-  \nu -\nu_1 - i \nu_2 =0$ when $\nu_1,\nu_2\to 0$, then the function $\Om\Lb\nu\,\vert\,\nu_1,\nu_2\Rb\he \sum^3_{i=1}\Om_i\Lb\nu\,\vert\,\nu_1,\nu_2\Rb$
tends to $-2\pi/3\nu_1\nu_2$, neglecting the contribution from $\Om_1$. Inserting this into \eq{planar} leads to the result:

\bea
&& \ml{\ml{\lim_{\substack{   i/2-  \nu -\nu_1 - i \nu_2 \to 0 \\ \nu_1,\nu_2\to 0   }}}}\hspace{0.3cm}
\Lb  i/2 - \nu  -\nu_1 -\nu_2\,\Rb\,\Ga_{\mbox{\footnotesize{planar}}}\Lb\nu\,\vert\, \nu_1,\nu_2\Rb\he
 \ml{\ml{\lim_{\substack{   i/2-  \nu \to 0 \\ \nu_1,\nu_2\to 0   }}}}\hspace{0.3cm}
\f{-2\,\pi\,\Lb\h+i\nu\Rb^2}{3\nu_1\nu_2}\lab{planartend}\eea

{\bf \large The non-planar vertex}

\vspace{0.3cm}

The formula for the triple Pomeron vertex of \eq{tpv} contained also the contribution of the non-planar diagram,
given by \eq{nonplanar}.
This expression included the function $\La \Lb\nu\vert\nu_1,\nu_2\Rb$
, which is given in terms of the following integral in ref. \cite{1.Korchemsky:1997fy}:

\bea
&&\La\Lb\nu\vert\nu_1,\nu_2\Rb=\!\!\Lint\f{d^2z_0d^2z_1}{\mid z_{01}\mid^4}\!\!
\Lb\f{z_{01}\bar{z_{01}}}{z_0z_1\bar{z}_0\bar{z}_1}\Rb^{\!\lh+i\nu}\!\!\!
\Lb\f{z_{01}}{\Lb 1-z_0\Rb\Lb 1-z_1\Rb}\f{\bar{z}_{01}}{\Lb 1-\bar{z}_0\Rb\Lb 1-\bar{z}_1\Rb}\Rb^{\lh+i\nu_1}\!\!
 \!\biggl( z_{01}\bar{z}_{01}\biggr)^{\lh+i\nu_2}\hspace{0.7cm}\lab{2.Lambda}
 \eea

The integrations of \eq{2.Lambda} were evaluated in ref. \cite{1.Korchemsky:1997fy}, where the following result was derived:

\bea
\La\Lb\nu\,\vert\,\nu_1,\nu_2\Rb\he2^{2i\nu+2i\nu_1+2i\nu_2-1}\,\pi^2\,\,\f{\Ga\biggl( \lh-i\nu\biggr) \Ga\biggl( \lh-i\nu_1\biggr) \Ga\biggl( \lh-i\nu_2\biggr) }
{\Ga\biggl( \lh+i\nu\biggr) \Ga\biggl( \lh+i\nu_1\biggr) \Ga\biggl( \lh+i\nu_2\biggr)}
\,\,
\hspace{5cm}\lab{lanu}\eea
$$\times \f{\Ga\biggl(\lf+\lh\Lb i\nu+i\nu_1+i\nu_2\Rb\biggr)}{\Ga\biggl(\mathlarger{\f{3}{4}}-\lh \Lb i\nu +i\nu_1+i\nu_2\Rb\biggr)}\f{\Ga\biggl(\lf+\lh\Lb i\nu+i\nu_1-i\nu_2\Rb\biggr)}
{\Ga\biggl(\mathlarger{\f{3}{4}}-\lh \Lb i\nu +i\nu_1-i\nu_2\Rb\biggr)}\,\,
\f{\Ga\biggl(\lf+\lh\Lb -i\nu+i\nu_1+i\nu_2\Rb\biggr)}{\Ga\biggl(\mathlarger{\f{3}{4}}-\lh \Lb- i\nu +i\nu_1+i\nu_2\Rb\biggr)}
\f{\Ga\biggl(\lf+\lh\Lb i\nu-i\nu_1+i\nu_2\Rb\biggr)}{\Ga\biggl(\mathlarger{\f{3}{4}}-\lh \Lb i\nu -i\nu_1+i\nu_2\Rb\biggr)}\nn$$

Using the above expression it is easy to see that:

\bea
&& \ml{\ml{\lim_{\substack{   i/2-  \nu -\nu_1 - i \nu_2 \to 0 \\ \nu_1,\nu_2\to 0   }}}}\hspace{0.3cm}
\Lb   i/2 - \nu  -\nu_1 -\nu_2\Rb\,\La\Lb\nu\,\vert\,\nu_1,\nu_2\Rb\he
 \ml{\ml{\lim_{\substack{   i/2-  \nu -\nu_1 - i \nu_2 \to 0 \\ \nu_1,\nu_2\to 0   }}}}\hspace{0.3cm}
\f{\pi^2}{2}\,\,\,\f{\Lb\h +i\nu\Rb}{\nu_1\nu_2}\lab{0.asymptoten}
\eea

Plugging this result into the definition of the non-planar diagram of \eq{nonplanar}, we
can derive  the non-planar diagram in the limit      
$  i/2-  \nu -\nu_1 - i \nu_2 \to 0$ where $\nu_1,\nu_2\to 0$.
Using the fact that $\lim_{i/2-\nu\to 0}\chi\Lb\nu\Rb\to 1/\Lb\h+i\nu\Rb$ (see the definition of \eq{chi}),
the asymptote is:

\bea
&& \ml{\ml{\lim_{\substack{   i/2-  \nu -\nu_1 - i \nu_2 \to 0 \\ \nu_1,\nu_2\to 0   }}}}\hspace{0.3cm}
\Lb   i/2 - \nu  -\nu_1 -\nu_2\Rb\,\Ga_{\mbox{\footnotesize{nonplanar}}}\Lb\nu\,\vert\,\nu_1,\nu_2\Rb\he
\ml{\ml{\lim_{\substack{   i/2-  \nu  \to 0 \\ \nu_1,\nu_2\to 0   }}}}\hspace{0.3cm}      
\f{\pi^2}{2}\,\,\,\f{\Lb\h +i\nu\Rb^2}{\nu_1\nu_2}\lab{nonplanartend}\eea

Finally, inserting the asymptotes of Eqs. (\ref{planartend}) and (\ref{nonplanartend}) into \eq{tpv},
we can calculate 
  $\tilde{\Ga}\Lb\nu\,\vert\,\nu_1,\nu_2\Rb\he
\Lb  i/2 - \nu  -\nu_1 -\nu_2 \Rb \Ga\Lb\nu\,\vert\,\nu_1,\nu_2\Rb$
in the region $  i/2-  \nu -\nu_1 - i \nu_2 \to 0$ where $\nu_1,\nu_2\to 0$  as:

\bea
&&
\ml{\ml{\lim_{\substack{ i/2 - \nu  -\nu_1 -\nu_2\to 0\\  \nu_1,\nu_2\to 0 }}}}\hspace{0.3cm}
\tilde{\Ga}\Lb\nu\,\vert\,\nu_1,\nu_2\Rb\he 
\ml{\ml{\lim_{\substack{ i/2 - \nu \to 0\\  \nu_1,\nu_2\to 0 }}}}\hspace{0.3cm}
\,\f{a\,\Lb   \h + i \nu\Rb ^2}{\nu_1\nu_2}
\lab{02.asymptotep}\\
\nn\\
\nn\\
&&\mbox{where}\hspace{0.5cm} a\he -\,\f{16\bas^2}{N_c}\,\,\,\,
\Lb \hspace{0.3cm} \f{2\pi}{3}\hspace{0.3cm}+\hspace{0.3cm} \f{\pi^3}{N_c^2}\hspace{0.3cm}\Rb\lab{definitionap}
\eea

In terms of the variables $\left\{\,\,\la,\si,\De\,\,\right\}$ defined in \eq{depm}, then \eq{02.asymptotep}
becomes:

\bea
&&
\ml{\ml{\lim_{\substack{ \la\to \si\\  \si\to 0 }}}}\hspace{0.3cm}
\tilde{\Ga}\Lb\la\,\vert\,\si,\De\Rb\he 
\ml{\ml{\lim_{\substack{ \la\to \si \\  \si\to 0 }}}}\hspace{0.3cm}\, 
\,\,\f{a\,\la ^2}{\Lb \la +\De \Rb\Lb \la - \De\Rb}
\lab{2.asymptotep}
\eea

\bibliographystyle{amsplain}
\label{sec:bib}

\end{document}